\documentclass[sigconf, nonacm, pdfa]{acmart}

\newcommand\vldbdoi{10.14778/3648160.3648166}
\newcommand\vldbpages{1227 - 1240}
\newcommand\vldbvolume{17}
\newcommand\vldbissue{6}
\newcommand\vldbyear{2024}
\newcommand\vldbauthors{\authors}
\newcommand\vldbtitle{\shorttitle} 
\newcommand\vldbavailabilityurl{https://github.com/jeongminpark417/GIDS}
\newcommand\vldbpagestyle{empty}

\usepackage{multirow}
\usepackage[T1]{fontenc}
\usepackage{newtxmath}
\usepackage{euscript}
\usepackage{listings}
\usepackage{ulem}
\usepackage{array}
\usepackage[a-2b]{pdfx}
\DeclareMathAlphabet{\mathpzc}{T1}{pzc}{m}{it}

\newcommand{\pname}[1]{{GIDS}{#1}}

\newcommand{\vm}[1]{{\textcolor{orange}{VM: \textit{#1}}}}

\usepackage{tikz}
\usepackage{xcolor}
\newcommand*\circled[1]{\tikz[baseline=(char.base)]{
            \node[shape=circle,fill,inner sep=0.5pt] (char) {\textcolor{white}{#1}};}}

\begin{document}
\title{GIDS: Accelerating Sampling and Aggregation Operations in GNN Frameworks with GPU Initiated Direct Storage Accesses}


\author{Jeongmin Brian Park}
\affiliation{%
  \institution{UIUC}\country{USA}
  }
\email{jpark346@illinois.edu}

\author{Vikram Sharma Mailthody}
\affiliation{%
  \institution{NVIDIA}\country{USA}
  }
\email{vmailthody@nvidia.com}

\author{Zaid Qureshi} 
\affiliation{%
  \institution{NVIDIA}\country{USA}
  }
\email{zqureshi@nvidia.com}

\author{Wen-mei Hwu}
\affiliation{%
  \institution{NVIDIA/UIUC}\country{USA}
  }
\email{whwu@nvidia.com}


\begin{abstract}
{

Graph Neural Networks (GNNs) are emerging as a powerful tool for learning from graph-structured data and performing sophisticated inference tasks in various application domains.
Although GNNs have been shown to be effective on modest-sized graphs, training them on large-scale graphs remains a significant challenge due to the lack of efficient  storage 
access and 
 caching methods for graph data.  
Existing frameworks for training GNNs use CPUs for graph sampling and feature aggregation, while the training and updating of model weights are executed on GPUs. 
However, our in-depth profiling shows CPUs cannot achieve the graph sampling and feature aggregation throughput required to keep up with GPUs.  
Furthermore, when the graph and its embeddings do not fit in the CPU memory, the overhead introduced by the operating system, say for handling page-faults, causes gross under-utilization of hardware and prolonged end-to-end execution time. 

To address these issues, we propose the GPU Initiated Direct Storage Access (\pname{}) dataloader, to enable GPU-oriented GNN training for large-scale graphs while efficiently utilizing all hardware resources, such as CPU memory, storage, and GPU memory.
The \pname{} dataloader first addresses memory capacity constraints by enabling GPU threads to directly fetch feature vectors from storage. Then, we introduce a set of innovative solutions, including the dynamic storage access accumulator, constant CPU buffer, and GPU software cache with window buffering, to balance resource utilization across the entire system for improved end-to-end training throughput. 
Our evaluation using a single GPU on terabyte-scale GNN datasets shows that the \pname{} dataloader accelerates the overall DGL GNN training pipeline by up to 582$\times$ when compared to the current, state-of-the-art DGL dataloader. 

}
\end{abstract}


\maketitle

\pagestyle{\vldbpagestyle}
\begingroup\small\noindent\raggedright\textbf{Reference Format:}\\
\vldbauthors. \vldbtitle. PVLDB, \vldbvolume(\vldbissue): \vldbpages, \vldbyear.\\
\href{https://doi.org/\vldbdoi}{doi:\vldbdoi}
\endgroup

\ifdefempty{\vldbavailabilityurl}{}{
\vspace{.3cm}
\begingroup\small\noindent\raggedright\textbf{Artifact Availability:}\\
The source code, data, and/or other artifacts have been made available at \url{\vldbavailabilityurl}.
\endgroup
}

\section{Introduction}

Owing to their expressive power, Graph Neural Networks (GNNs) effectively capture the rich relational information embedded among input nodes and edges, leading to improved generalization performance over traditional machine learning techniques. 
As a result, GNNs have gained significant attention in recent years and demonstrated their efficacy in 
graph-based machine learning applications, such as node classification~\cite{GCN,Graphsage,GAT,LazyGCN}, recommendation~\cite{gnn-recommendation, pinner_sage}, fraud detection~\cite{gnn_fraud,fdgar,gnn_fraud3,gnn_fraud4}, and link prediction~\cite{gnn_linkpredict,fewshot,LP_system}.

To cater to this growing interest, new open-source frameworks such as PyTorch Geometric (PyG)~\cite{pyg}, Spektral~\cite{Spektral}, and Deep Graph Library (DGL)~\cite{dgl} have been developed to provide optimized operators required by GNNs, such as message-passing for aggregating feature information across related graph nodes, and graph-specific neural network computation layers. Although GNN frameworks leverage {GPUs' high-throughput tensor operations}, 
GNN training faces challenges beyond {its} computational requirement. 
A major challenge is the fast-growing graph dataset sizes that cannot fit into the limited GPU memory capacity. To address this challenge, frameworks like DGL exploit Unified Virtual Addressing (UVA) by pinning both the graph structure data and feature data into the CPU memory, thus enabling GPU kernels to efficiently perform subgraph extraction and feature aggregation while making zero-copy access to the graph data from the CPU memory~\cite{pytorch-direct}. 

For large-scale graphs that do not fit into the CPU memory, the UVA approach is no longer sufficient. There are classes of traditional solutions to support large-scale GNN training: (a) multi-node/multi-GPU, (b) tiling, and (c) memory-mapped files. Leveraging multiple nodes or GPUs~\cite{DSP, star, ROC, Neugraph, mg_gcn} 
 by partitioning the graph across the nodes/GPUs to support large-scale GNN training is an expensive approach~\cite{scaling_out}. 
Tiling~\cite{featgraph,zipper} can be used to support large-scale GNN training by leveraging graph partitioning to move tiles of graph data  
in and out of the GPU memory. This approach shows poor performance due to random access patterns and the additional cost of pre-processing the input data. 
Finally, the most convenient solution to train large-scale graph datasets on a single GPU is exploiting the memory-mapped file technique, which maps the graph data stored on disk to the GPU's virtual address space, 
enabling the GPU to access 
the data without first loading the entire dataset into memory. 
Previous studies~\cite{Ginex, AliGraph, PaGraph, zero} extended the memory-mapped file approach and leveraged the in-memory caching mechanism to mitigate the storage access overhead. 

Despite its conceptual simplicity, the use of memory-mapped files in GNN training faces performance challenges due to the heavy software overhead in handling page faults and its inability to tolerate long latency incurred during data retrieval from storage. The storage latency, which is two to three orders of magnitude longer than the DRAM access latency, becomes a bottleneck in the GNN training process. This is due to sparse and irregular graph data access patterns and the inability of the memory-mapped file approach to overlap the latencies of these accesses, resulting in poor overall performance. 
In Section 2.3, we show that when using memory-mapped files, the sampling and feature aggregation stages of the GNN training pipeline dominate the overall execution time and severely limit the overall GNN training performance.

In this paper, we propose a new approach called GPU Initiated Direct Storage Access (\pname{}) dataloader to tackle the challenges of GNN training on large-scale graphs by leveraging GNN-specific characteristics to efficiently utilize all the involved hardware resources (CPU memory, storage, and GPU memory). 

\begin{figure}[h]
    \centering
    \includegraphics[width=\columnwidth]{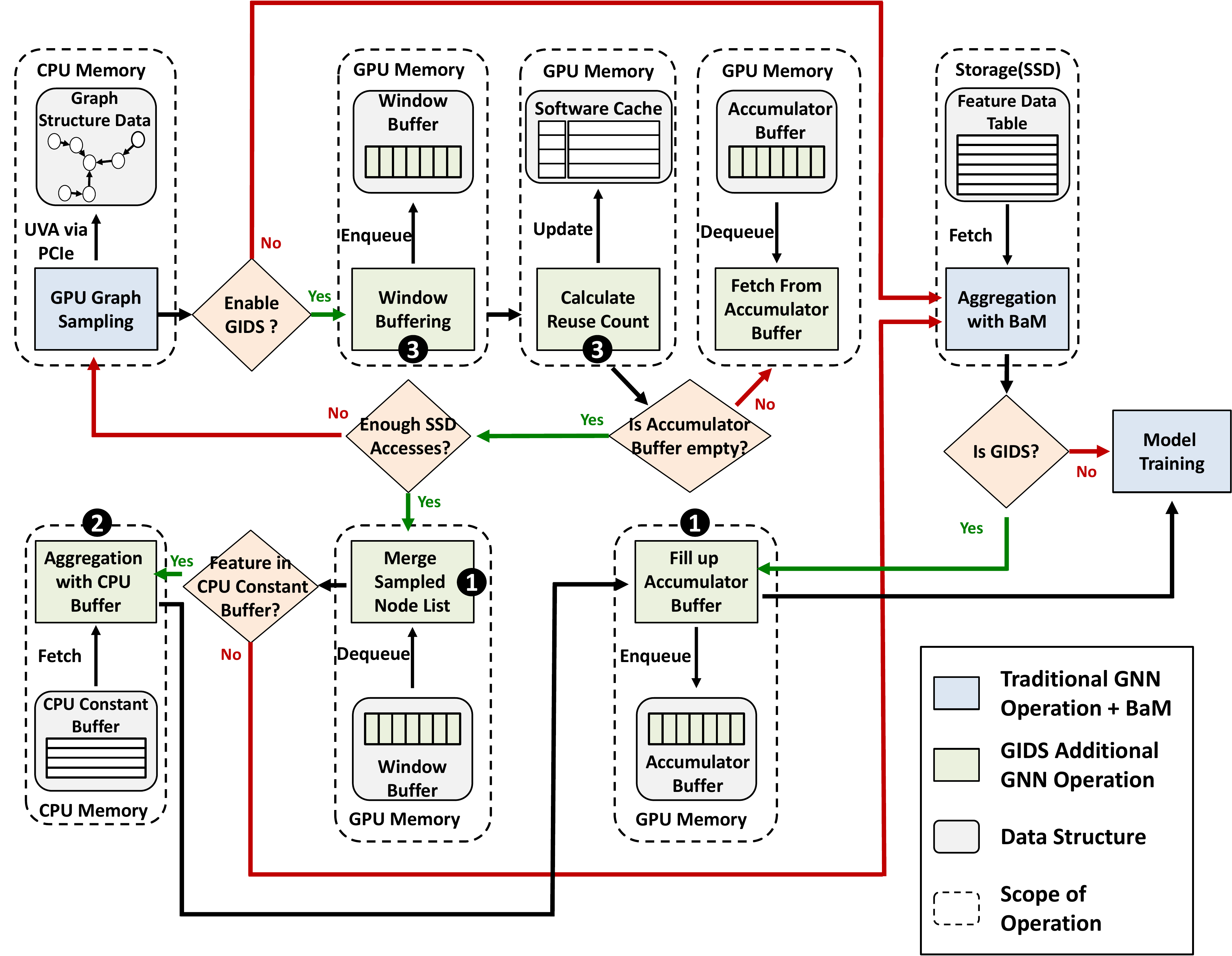}
    \caption{ Illustration of the GNN training process with the \pname{} dataloader or the BaM dataloader. 
   }
    \label{fig:workflow}
\end{figure}

Figure~\ref{fig:workflow} illustrates the GNN training workflow with the \pname{} dataloader.
First, \pname{} keeps the feature data of the graph in storage as the feature data typically accounts for the vast majority of the total graph dataset size for large-scale graphs (see Table~\ref{tab:dataset_percent} for details). 
\pname{} overcomes the long storage access latency by leveraging BaM~\cite{bam} to allow GPU threads to directly fetch feature data, using the massive GPU thread-level parallelism to overlap the latencies of many storage accesses. 
However, to achieve the peak SSD bandwidth, ensuring a sufficient number of concurrent storage access requests is a critical prerequisite. 
The number of storage access requests can vary based on the sampling parameters or hardware configuration.
To maintain sufficient overlapping storage accesses for any environment, \pname{} features a \textit{dynamic storage access accumulator} (1), a novel technique that exploits the independency of the graph sampling process to automatically merge iterations based on the system hardware specification.

Second, \pname{} pins the graph structure data, whose size is typically tiny compared to the feature data, in the CPU memory to enable GPU graph sampling via UVA zero-data copy transfer to avoid I/O amplification and cache pollution. 
Third, \pname{} enables users to 
 reserve CPU memory for a \textit{constant CPU buffer} (2) to achieve higher feature aggregation effective bandwidth by redirecting accesses from storage to the constant CPU buffer for hot nodes when PCIe bandwidth is not fully utilized. 
 
Finally, \pname{} allocates GPU memory for the BaM Application-Defined Software Cache to store feature data for recently accessed nodes to minimize the storage accesses. As a new contribution to the cache design, we introduce a novel \textit{window buffering} (3) technique that takes advantage of the timing flexibility of the graph sampling process to exploit locality across mini-batches and further improve GPU cache utilization.

We make the following key contributions in this paper.
\begin{itemize}
\item We analyze the limitations of the existing GNN frameworks while training on large graph datasets and show that the existing CPU-initiated approach cannot keep up with the demands of GPU-accelerated GNN training.


\item We introduce a novel dynamic storage access accumulator, which accurately estimates the required number of overlapping storage access requests to achieve peak SSD bandwidth with the BaM system. The accumulator automatically maintains a sufficient number of storage requests by decoupling the graph sampling stage from the training stage and allowing the former to run ahead of the latter.

\item We conduct an analysis of the logical dependencies among 
the overall GNN training pipeline stages and propose an innovative optimization strategy named window buffering. It enables the \pname{} dataloader to foresee the upcoming node access pattern to optimize the cache eviction policy.

\item We increase the effective bandwidth of the \pname{} dataloader beyond the limited SSD bandwidth by classifying hot nodes using reverse page rank scores, storing them in the constant CPU buffer, and redirecting some of the storage accesses to the CPU memory when PCIe bandwidth is underutilized due to limited SSD bandwidth. 

\end{itemize}

We demonstrate the \pname{} dataloader's effectiveness and flexibility by measuring performance using billion-scale datasets that do not fit in the CPU memory. 
The results based on the NVIDIA A100 GPUs and 512GB CPU memory capacity show that the \pname{} dataloader achieves 582$\times$ speedup in overall training over the state-of-the-art GNN dataloader.






\section{Background}
\label{sec:background}
In this section, we provide an overview of GNN models, followed by an introduction to mini-batching and sampling-based GNN training. We then explain the state-of-the-art framework for large-scale GNN training and its challenges.

\subsection{Graph Neural Networks (GNNs)}

Graph Neural Networks (GNNs) have recently gained prominence in solving machine learning problems by incorporating graph structure information~\cite{GCN, GAT,gcn2,gnn2}. 
These networks typically consist of multiple layers and operate through layer-wise message passing.

Given a graph $\mathcal{G}(\mathcal{V},\mathcal{E})$, with vertex set $\mathcal{V}$ and edge set $\mathcal{E}$, the node feature vectors for each vertex $v \in \mathcal{V}$ are represented as $x_v$. The node embedding of vertex $v$ at layer $l$ is denoted as $h_v^{(l)}$, with $h_v^{(0)}$ initialized with $v$'s feature vector $x_v$. The GNN updates the node embeddings using the equation:
\begin{equation}
h_v^{(l+1)} = f(h_v^{(l)},{h_w^{(l)}}_{w \in \mathcal{N}(v)}),
\end{equation}
where $\mathcal{N}(v)$ defines the neighborhood set of $v$, $h_w^{(l)}$ denotes the node embedding of the neighbor node $w$ at layer $l$, and $f$ is a parameterized update function.

Graph data consists of two components: graph structure data and node feature data. The graph structure data represents the edges and nodes of the graph, while the node feature data represents the feature embeddings for each node. Sparse matrix formats such as Coordinate (COO) format and Compressed Sparse Column (CSC) format are commonly used to store the graph structure data, whereas the node features are typically stored in an $N \times D$ matrix, where $N$ is the total number of nodes in the graph, and $D$ is the dimension of each node feature. 
The size of each node's feature can vary greatly but typically ranges from 512B to 4KB. 
For large-scale graphs with billions of nodes, the size of the node feature data can reach several tens of terabytes. 
As a result, managing the node feature data for large-scale GNN training with limited memory capacity is a challenging task.

\subsection{GNN Training Pipeline}
GNN training on large graph datasets involves mainly four stages: graph sampling, feature aggregation, data transfer, and model training. 
Mini-batch training is commonly used in these models for scalability and computational efficiency~\cite{minibatch1, minibatch2, minibatch3}. In this section, we briefly describe the mini-batching technique and each key stage of the GNN training pipeline.

\subsubsection{Mini-batching}
Mini-batching of GNN models involves splitting the graph into smaller sub-graphs and training the network on each of these sub-graphs. 
    During each iteration of the training process, a batch of sub-graphs is loaded into GPU memory for computation. 
The batch size must be carefully chosen to prevent GPU memory overflow during training. 
Mini-batching also exposes more parallelism as mini-batches can be assigned to different GPUs during training,
which significantly improves training speed and efficiency and makes it a popular approach for many GNN models.
Previous studies have demonstrated that training neural networks with 
mini-batches can also lead to faster convergence and better optimization compared to training on the entire dataset~\cite{minibatch1, minibatch2,minibatch3}.

\subsubsection{Node Sampling}
\begin{figure}[ht]
    \centering
    \includegraphics[width=\linewidth]{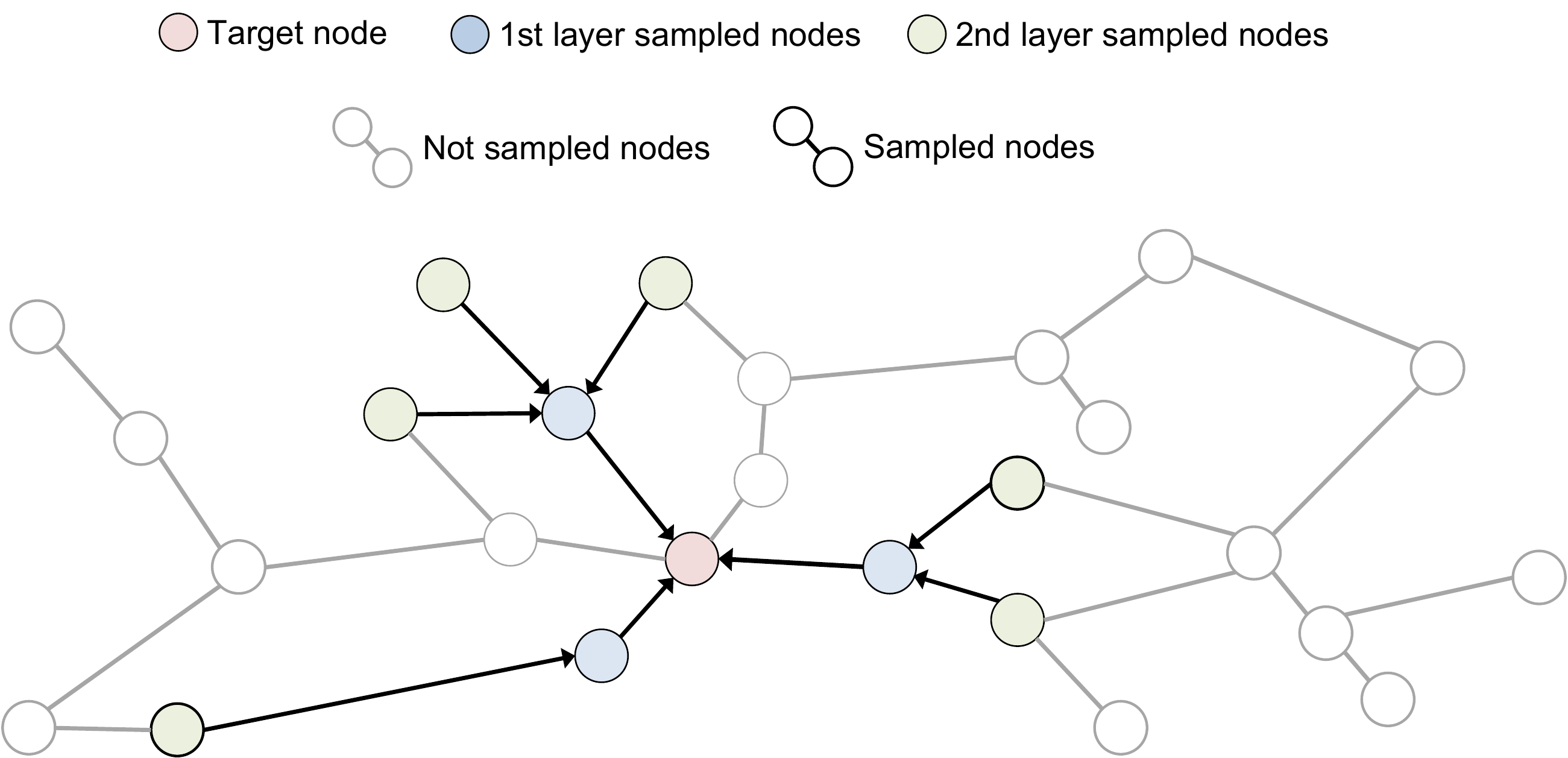}
    \caption{A subgraph generated by a uniformly random selection method for two-layer Neighborhood Sampling. 
    }
    \label{fig:sampling}
\end{figure}

Mini-batching alone cannot fully address the scalability limitations when working with large graphs. 
Even with small batch sizes, the training cost can still be substantial due to the exponential growth of memory footprint when collecting k-hop neighbors. 
GraphSAGE~\cite{Graphsage} introduced the concept of neighborhood sampling to tackle this problem. 
GraphSAGE reduces the computation and memory footprint by randomly sampling a fixed number of neighboring nodes rather than including all nodes in the graph. 
To ensure a sufficient level of randomness in the training process, GraphSAGE uses a uniformly random selection method for neighborhood sampling. 
Figure ~\ref{fig:sampling} illustrates an example of neighborhood sampling with a 2-hop computational graph. 
In this example, the sampling size is set to 3, meaning up to three neighboring nodes of the target node are selected. 
With two layers, the sampled subgraph consists of 10 (1 + 3 + 6) vertices and 9 edges.

\subsubsection{Node Feature Aggregation} 
The node features for 
the sampled subgraph of a mini-batch must be aggregated, or gathered, before 
training on the mini-batch can start. For smaller graphs whose node feature data can fit into the CPU memory, the entire feature data is first loaded into the CPU memory. The node features for each mini-batch's sampled subgraph are gathered from the CPU memory and transferred into the GPU memory. 
In cases where node feature data for the original graph exceeds the CPU memory capacity, the current state-of-the-art approach~\cite{dgl, IGB}  uses the CPU to first gather the node features of the sampled subgraph from storage into a buffer in the CPU memory, and then transfer the buffered feature data from the CPU memory to the GPU memory. 

\subsection{Limitation of Existing GNN Frameworks} 
\label{sec:motivation}

\begin{figure}[h]
    \centering
    \includegraphics[width=\columnwidth]{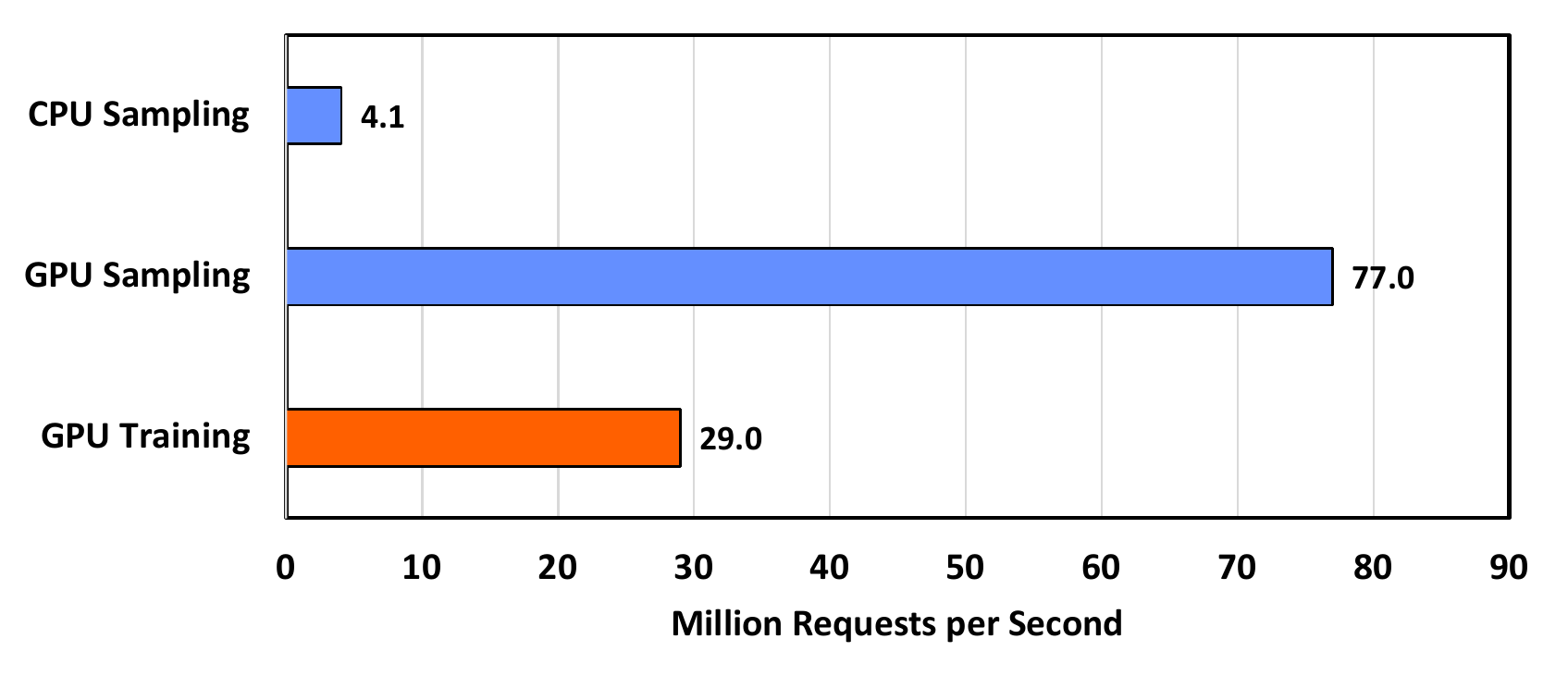}
    \caption{Request generation rate of data preparation on CPU and GPU, and request consumption rate on GPU on IGB-small dataset. The CPU and GPU used in this measurement are listed in Table~\ref{tab:config}.} 
    \label{fig:cpu_io} 
\end{figure}

State-of-the-art GNN frameworks, such as DGL~\cite{dgl} and PyG~\cite{pyg}, 
have adopted a hybrid CPU-GPU training system, 
where the CPU is responsible for data preparation, and the GPU handles the model training. 
 Our profiling results show that such a hybrid training approach can lead to significant under-utilization of the GPU and suboptimal training time.
Figure~\ref{fig:cpu_io} 
compares the node feature vector request generation rate 
of the data preparation stages,  i.e., node sampling and node feature aggregation, of the GNN 
training pipeline when these stages are executed on the CPU vs. on the GPU.
As a reference, Figure~\ref{fig:cpu_io} also shows that the training kernels running on the GPU can consume 
the aggregated node features at a rate
of 29 million requests per second. 
To maximize GPU utilization and minimize GNN training time for large graphs, the 
request generation rate must match or exceed the consumption rate.

However,  as shown in Figure~\ref{fig:cpu_io},  the data preparation stages cannot generate more than 4.1 million feature vector requests per second, 
even when using multiple threads (16 in this experiment beyond which the rate plateaus) on the CPU.
This is because the sampling computation involves repeatedly traversing the graph and accessing its edges and nodes, 
making it difficult for the CPU, with its limited memory bandwidth and thread-level parallelism, to keep up with the consumption rate of the GPU-accelerated training kernels. 
In contrast, the GPU can generate 77 million feature requests per second, which is more than sufficient to match the consumption rate of the training kernels. 
Based on these observations, we will focus on GNN training pipelines that offload the data preparation stages to the GPU for the remainder of the paper.

A challenge in running the data preparation stages on the GPU is the limited GPU memory capacity that can be significantly smaller than the CPU memory.
To address this challenge, DGL recently introduced the UVA-based GNN training technique~\cite{pytorch-direct}, which pins the entire graph dataset (both graph and feature vectors) in the CPU memory and enables the graph sampling and feature aggregation kernels running on the GPU to directly access the graph dataset through zero-copy accesses. 
While this approach helps to scale GNN training to 
graph datasets 
whose sizes exceed the GPU memory capacity, it cannot handle large-scale graphs whose sizes 
surpass the capacity of the CPU memory since all graph data must be pinned in the CPU memory for the UVA-based technique to work.

\begin{figure}[h]
    \centering
    \includegraphics[width=\columnwidth]{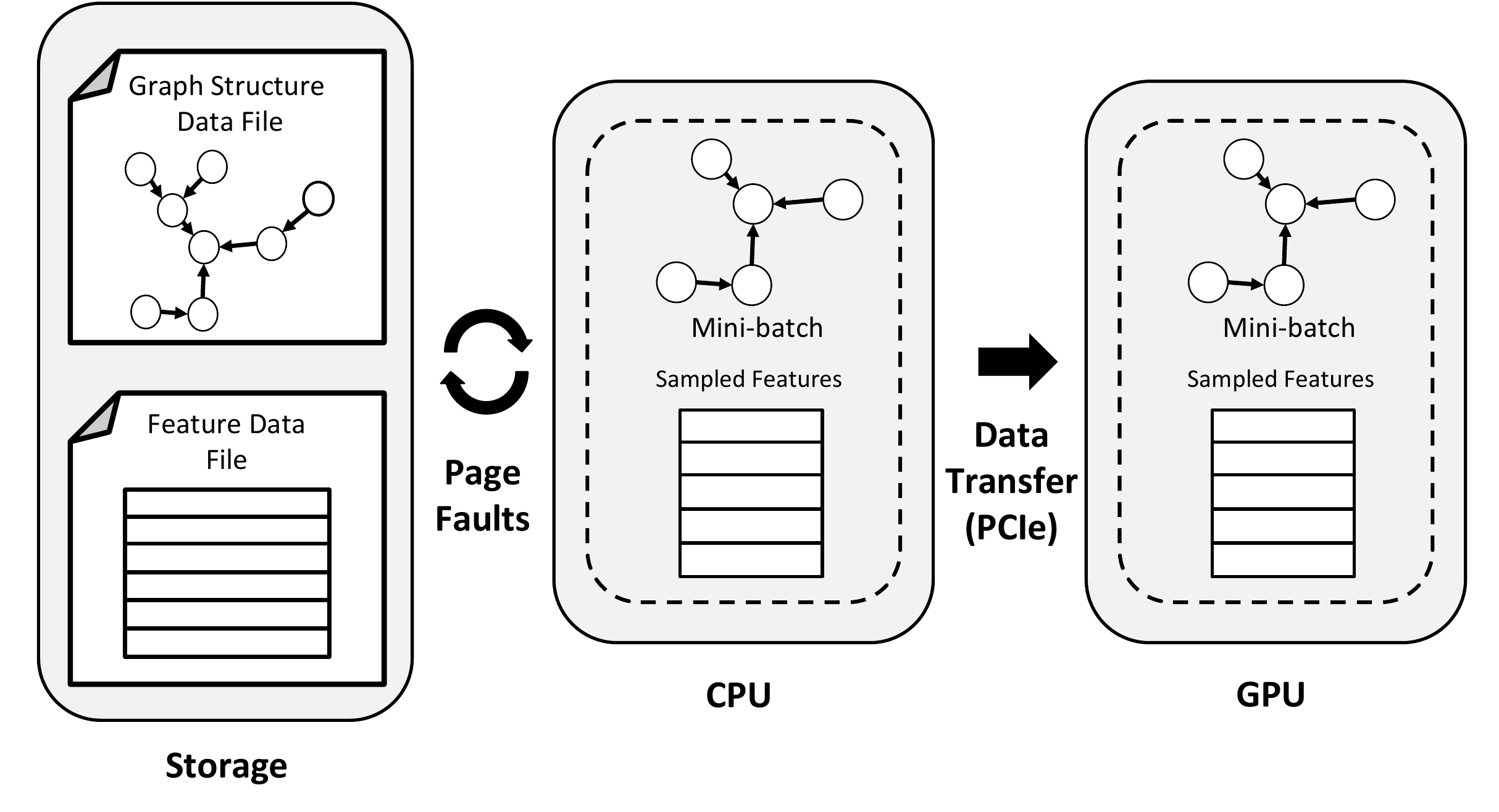}
    \caption{Illustration of the GNN training process with the memory-mapping DGL dataloader}
    \label{fig:cpu_workflow}
\end{figure}

Existing GNN frameworks fall back to the CPU for graph sampling and feature aggregation execution to support graph datasets that cannot fit into the CPU memory.
The key idea is to provide a notion of infinite virtual memory by memory-mapping 
the node feature vector files into the CPU virtual address space 
and allow the node feature aggregation computation on the CPU to page fault when the requested feature vector is unavailable in the CPU memory. Figure~\ref{fig:cpu_workflow} illustrates the GNN training process using the approach of the memory-mapped file in the DGL framework. During the node feature aggregation stage, the CPU accesses the node features mapped in its virtual memory space, and the OS page fault handler brings 
the pages that contain the accessed features from storage into the CPU memory when it misses from the OS page cache. 
The memory-mapped file approach, along with the CPU execution of node feature aggregation, eliminates the need for loading/pinning the entire dataset into the CPU memory \textit{a priori} and only brings in the data being actively used on-demand. 


\begin{figure}[h]
    \centering
    \includegraphics[width=\columnwidth]{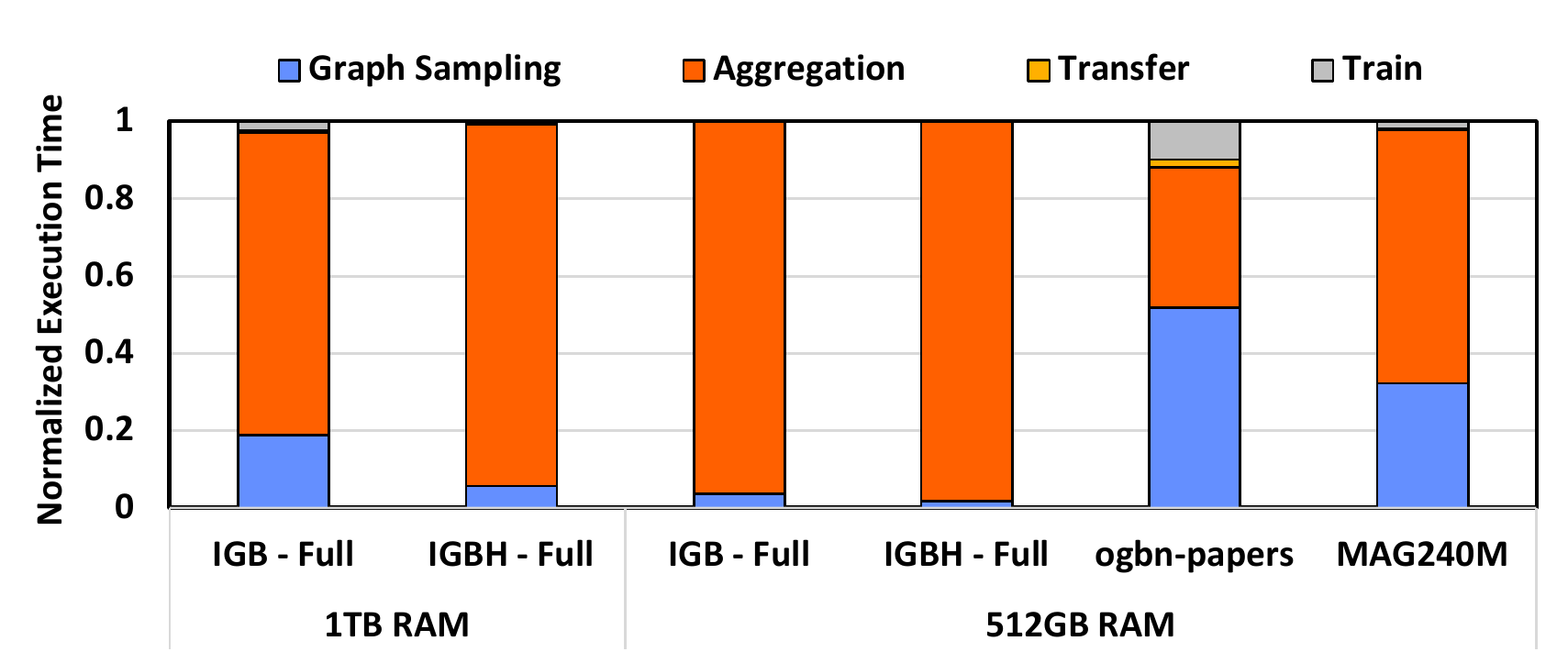}
    \caption{GNN training time breakdown for the baseline DGL dataloader for different graph datasets. The node feature data is accessed from memory-mapped files, while the graph structure data is stored in the CPU memory. The GraphSAGE model is used as the GNN training model. The graph properties are listed in Table~\ref{tab:dataset}.}
    \label{fig:breakdown}
\end{figure}

Unfortunately,  the memory-mapped file approach makes the node feature aggregation by far the worst bottleneck of the overall training pipeline.
Our profiling of each stage in the GNN training execution shows the iteration time is 
clearly dominated by the sampling and node aggregation stages, as shown in Figure~\ref{fig:breakdown}. 
For example, the training stage is barely visible for the IGB-Full and IGBH-Full graphs, the largest two graphs used in our evaluations.
This is because, for large-scale graphs, the additional cost of page faults exacerbates the gap between 
the data preparation throughput and model training throughput. 
\textit{Thus, the key to improving the GNN training performance while training on large graphs 
is to drastically accelerate the sampling and feature aggregation stages (i.e., the data preparation stages). }

Previous research \cite{PaGraph, AliGraph, Ginex, gnn_pipe} has aimed to enhance the efficiency of node aggregation and sampling stages running on the CPUs by using specific in-memory caching mechanisms to minimize redundant storage accesses and/or utilizing pipelining techniques to conceal graph sampling time. 
However, as shown in Figure~\ref{fig:cpu_io}, the data preparation stages running on the CPU cannot even generate node feature requests at a sufficiently high rate to match the consumption rate of the training kernels. A successful solution to the problem of efficiently accessing node feature vectors from the storage on-demand must allow the data preparation stages running on the GPU to make direct requests to the storage devices.

To this end, we develop the GIDS dataloader based on the BaM~\cite{bam} software stack, a recently released research infrastructure that enables direct storage device access by the GPU, eliminating the overhead of OS page faults during feature vector data access.

\subsection{The BaM System}
The BaM system~\cite{bam} aims to tackle the problem of storage latency in big-data GPU applications. The key idea behind BaM is to allow GPU threads to have direct access to the storage. 
As a massive number of GPU threads can initiate direct storage access without incurring CPU-GPU synchronization or CPU software overhead, the GPU can take full advantage of parallelism to hide long storage access latency, enabling it to achieve peak storage bandwidth 
 when there is a sufficient number of concurrent storage access requests.

However, straightforward adoption of BaM in the data preparation stages of the GNN training pipeline leaves much end-to-end performance on the table due to {the} imbalanced use of critical resources in the system. Therefore, we propose a suite of novel techniques to shift the use of hardware resources during the data preparation stages and significantly improve the end-to-end GNN training time.

\section{System Design}

To address the challenges associated with state-of-the-art large-scale GNN training, we design and implement the \pname{} dataloader, which enables fully GPU-oriented GNN training for large graphs and efficiently utilizes hardware resources.
This section describes the design and optimization of the \pname{} dataloader\footnote{Although the GIDS discussion in this section is based on DGL framework, it can be easily extended to other GNN frameworks such as PyG~\cite{pyg} and AliGraph~\cite{AliGraph}.}.

\subsection{ {\pname{} Dataloader System Overview}}

The \pname{} dataloader improves the performance and scalability of GNNs by efficiently utilizing all available hardware resources when aggregating node features that cannot fit into the CPU memory.
This section provides a detailed breakdown of how each resource is harnessed to accelerate GNN data preparation.
The illustration of the GNN training workflow with the \pname{} dataloader is shown in Figure~\ref{fig:workflow}.

\textbf{GPU: } As discussed in Section~\ref{sec:motivation}, the \pname{} dataloader moves the data preparation stages from the CPU to the GPU. As shown in Figure 3, the request generation rate of the sampling and node feature aggregation stages running on the GPU exceeds the GPU training kernel throughput.

\textbf{Storage: } To overcome memory capacity constraints, the \pname{} dataloader stores the feature data in storage. To address the challenge of storage access bottlenecks, the \pname{} dataloader employs the BaM system~\cite{bam}, enabling GPU threads to directly access storage and bypassing CPU page-fault handling software overhead.

The \pname{} dataloader also features a novel {\bf{dynamic storage access accumulator}} to merge iterations for the graph sampling and feature aggregation processes, ensuring a sufficient number of concurrent storage accesses (Section~\ref{sec:ssd}).

\textbf{CPU Memory: } The \pname{} dataloader pins the graph structure data in CPU memory because the graph structure data (4-8B) accessed by the sampling process exhibits a much finer granularity access pattern than the node feature data (512-4096B) accessed by the aggregation process. Despite the graph structure data being pinned in memory, this does not lead to memory capacity issues as it constitutes only a small fraction, typically around 5\% of the total dataset size. The structure data comfortably fits within CPU memory, even for terabyte-scale graphs (refer to Table~\ref{tab:dataset_percent}).

The \pname{} dataloader allocates a user-configurable portion of CPU memory as a {\bf{constant buffer}} to pin a small subset of the node feature data.
This buffer redirects storage accesses for hot nodes to the constant CPU buffer, amplifying the effective feature aggregation bandwidth beyond the available SSD bandwidth (Section~\ref{sec:feature_buffer}).

\textbf{GPU Memory: } The \pname{} dataloader employs BaM's application-defined software cache to temporarily store the feature data of recently accessed nodes in the GPU memory. This reduces the number of storage accesses and improves feature aggregation performance. Additionally, the \pname{} dataloader runs the data preparation several iterations ahead of the training stage and maintains a node access list for future iterations in a {\bf{window buffer}}, enhancing the GPU software cache hit ratio by leveraging GNN-specific data access patterns (Section~\ref{sec:window_buffering}).


\subsection{Dynamic Storage Access Accumulator}
\label{sec:ssd}

The \pname{} dataloader leverages the BaM system and takes advantage of the massive thread-level parallelism provided by GPUs to effectively handle storage latency during feature aggregation. To achieve this, a critical prerequisite is ensuring a sufficient number of concurrent storage access requests during the feature aggregation stage to maximize the utilization of the peak storage throughput.

The feature aggregation kernel via the BaM system can be divided into three distinct stages. The first stage is the initial stage, occurring from the beginning of feature aggregation until the first data is fetched from the SSD. The second stage is the steady-state stage, where data reception from the SSD reaches its peak IOPs. The final stage is the termination stage, the time between when the last access request to the SSD is handled and the conclusion of the feature aggregation process. During the initial and termination stages, SSD bandwidth utilization is almost zero, while it reaches its peak during the steady-state stage. Using this information, one can calculate the number of overlapping storage access requests required to achieve peak SSD's read throughput based on the following mathematical equations: 

\begin{equation}
    \label{eq:SSD_bw}
    N_{access} = IOP_{achieved} * (T_{i} + T_{s} + T_{t}) *  N_{ssd} 
\end{equation}
\begin{equation}
    \label{eq:SSD_time}
    T_s = \frac{N_{access}}{IOP_{peak}}  
\end{equation}

where $N_{access}$ defines the required number of concurrent storage accesses that must be maintained over time. 
$T_{i}$, $T_{s}$, and $T_{t}$ denote the time spent during the initial, steady-state, and termination stages, respectively. 
$IOP_{peak}$ represents the peak IOPs for each SSD while $IOP_{achieved}$ is the average achieved IOPs per SSD during the feature aggregation stage.
Finally, $N_{ssd}$ is the number of SSDs connected to a single GPU.

In general, one determines the $N_{access}$ value by making $T_s$ much larger than $T_i + T_t$, which can be determined empirically for the system used. Naively, one can increase $T_s$ by increasing the mini-batch size.
While the size of the mini-batch can be adjusted based on available computational resources and task-specific requirements, it may not be able to increase beyond a certain point due to training quality considerations. Thus, one needs to eliminate the stop-and-go boundaries between mini-batches by merging the data-preparation of consecutive iterations and thus effectively increase $T_s$. Statically setting the number of iterations to merge to effectively hide storage latency with the BaM system is not straightforward. 
Setting the numbers too small will result in poor $IOP_{achieved}$ and setting the number too high will incur an excessive level of buffer memory usage.

The required number of concurrent storage accesses depends on the characteristics of the SSD, with SSDs exhibiting higher latency $T_i$ demanding even more concurrent accesses. Furthermore, if multiple SSDs are connected to increase the collective SSD bandwidth, the required number of concurrent storage accesses scales linearly with the number of SSDs. 
Additionally, the \pname{} dataloader leverages both CPU and GPU memory to amplify feature aggregation throughput by redirecting some storage accesses to CPU/GPU memory. Thus, ensuring an adequate number of node feature accesses is crucial, as there must be enough storage access availability even after some accesses are redirected.

To address this challenge, we introduce the dynamic storage access accumulator within the \pname{} dataloader. 
This innovative approach takes advantage of the work independence inherent in the graph sampling process. 
Notably, the graph sampling and feature aggregation stages of an iteration are logically independent of the model training stage of previous iterations because the output of the model training stage solely updates model parameters and does not impact graph sampling or feature aggregation of future iterations. 

Based on this observation, the dynamic storage access accumulator combines iterations for the graph sampling and feature aggregation processes to maintain a sufficient number of concurrent storage accesses over time. Initially, it calculates the threshold for the required node accesses based on the proposed mathematical model.
The accumulator executes graph sampling processes for future iterations until the number of node accesses surpasses the threshold. 
At this point, the \pname{} dataloader enters the steady state, retrieving node feature vectors into mini-batch buffers in the GPU memory and starting new iterations as the accesses for the older iterations are completed.  
The training stage makes progress by accessing the next mini-batch from the batch buffers and performing model training on the mini-batch.

Note that the number of required node accesses is influenced by the number of storage accesses redirected to CPU/GPU memory. Therefore, the dynamic storage access accumulator tracks the number of redirected storage accesses and dynamically adjusts the threshold value accordingly.

\subsection{Constant CPU Buffer} 
\label{sec:feature_buffer}

By exploiting the GPU's massive parallelism and a sufficient number of concurrent storage requests, the \pname{} dataloader can achieve peak SSD read bandwidth during feature aggregation. However, it is crucial to acknowledge that the peak read bandwidth of a single SSD falls significantly short of the PCIe bandwidth, which is typically around 32GB/s. 
For instance, the peak read IOPs for Intel Optane SSDs is around 1.5 million requests per second with a 4KB cache-line granularity (equivalent to 6GB/s)~\cite{bam, Optane}, whereas NAND Flash SSDs can only reach a maximum of 800  thousand requests per second (approximately 3.2GB/s). 

BaM~\cite{bam} addresses this challenge by connecting multiple SSDs to a single GPU, thereby linearly scaling the collective SSD bandwidth to saturate the PCIe bandwidth. 
However, implementing such a system may not be practical for GNN developers, as it typically requires the connection of at least 4 to 5 Intel Optane SSDs to a single GPU. In the case of Samsung 980pro SSDs, a more substantial number, exceeding 10 SSDs or more, may be required to fully saturate the PCIe bandwidth. 

To amplify the effective bandwidth of the feature aggregation process when the PCIe bandwidth is under-utilized, the \pname{} dataloader leverages CPU memory as a constant CPU buffer. 
When there is available CPU memory, \pname{} offers users the flexibility to allocate a configurable portion of CPU memory as a constant CPU buffer to pin a portion of the feature data into the CPU memory.

Accesses to the feature table in the SSD are redirected to the constant CPU buffer when the requested feature data resides in the constant CPU buffer. With the assistance of the \pname{} dataloader's storage access accumulator, there remains a sufficient number of storage accesses to hide the storage latency, thereby preserving peak SSD read bandwidth.
As these redirected accesses are managed by CPU memory, the effective bandwidth of the feature aggregation process increases proportionally to the number of redirected accesses until the GPU ingress PCIe bandwidth is fully utilized. 

To maximize the utilization of the constant CPU buffer, it is essential to optimize the number of redirected accesses. This can be achieved by leveraging the access pattern of the graph sampling process. Prior research~\cite{DataTiering} demonstrates the use of weighted reverse page rank as a metric for distinguishing between hot nodes and cold nodes can help in this regard. As a result, the \pname{} dataloader selectively retains nodes with the highest weighted reverse page rank in the constant CPU buffer. Furthermore, the \pname{} data loader provides users with the flexibility to define which nodes should be pinned in the static GPU buffer when alternative metrics are more suitable for identifying hot nodes.

\subsection{Window Buffering} 
\label{sec:window_buffering}

Although the \pname{} dataloader can effectively handle storage latency and achieve up to PCIe bandwidth during feature aggregation, the achievable storage read bandwidth is orders of magnitude lower than the GPU memory bandwidth as High Bandwidth Memory 2 (HBM2) of recent NVIDIA GPUs can provide ~2TB/s bandwidth~\cite{HBM2} whereas the storage read bandwidth is limited by the 32GB/s PCIe in-take bandwidth of A100. Therefore, efficient utilization of GPU memory is necessary to amplify the effective bandwidth and further accelerate the feature aggregation process.

To address the bandwidth shortfall, \pname{} employs BaM's GPU application-defined software cache.
Unlike the GPU hardware caches, which help to conserve DRAM bandwidth, the \pname{} software cache is used to help conserve storage bandwidth. 
BaM's software cache temporarily stores the previously accessed cache-lines based on the random eviction policy.

However, when the graph dataset is much larger than the GPU cache, achieving high reusability of node feature data becomes challenging due to the random nature of the neighborhood sampling process. In such scenarios, it is critical to accurately predict the cache-lines that will be reused in the near future.

\begin{figure*}[h]
    \centering
    \includegraphics[width=\textwidth]{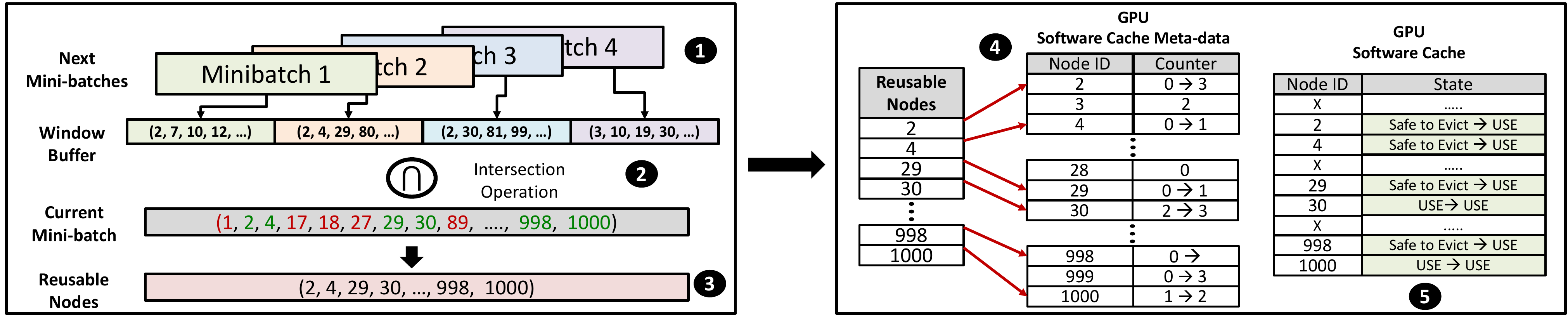}
    \caption{Example of the window buffering technique from  \pname{} dataloader}
    \label{fig:window_buffering}
\end{figure*}

To overcome this challenge, the \pname{} dataloader introduces a novel technique called window buffering and integrates it into the software cache. Unlike the traditional frameworks, \pname{} leverages the BaM software-defined cache which supports the customization of cache-line eviction policies. 
The window buffering technique reduces cache thrashing by avoiding the eviction of reusable node feature vectors through mini-batch look-ahead. 
This is achieved by conducting a graph sampling operation for a configurable number of iterations to fill the window buffer with sampled node IDs and avoiding the eviction of feature vectors for reused nodes in the window buffer.
Therefore, the dataloader can look-ahead to the list of the sampled nodes for 
future iterations.

Specifically, as illustrated in Figure~\ref{fig:window_buffering}, the window buffer in the \pname{} dataloader is initially filled with the node IDs that will be sampled in the next few iterations (\circled{1}). Once the window buffer is filled, the sampled node IDs in the current mini-batch are compared with the nodes in the window buffer (\circled{2}). Then, the list of nodes that will be reused in the next iterations and the number of occurrences is generated (\circled{3}). This information is then used to update the software cache metadata, which tracks the number of reuses in the next iterations for each node (\circled{4}).

During the update, when the future reuse counter value changes from 0 to any positive number, the state of the node in the GPU cache is changed from the “Safe to Evict” state to the "USE" state so that the corresponding cache-line will not be evicted. If the counter value is already a positive number, the state is kept marked as the "USE" state (\circled{5}).
The counter value is decreased each time the node is reused during the feature aggregation stage. 
When the counter value becomes 0, the state of the corresponding cache-line is then set back to the “Safe to Evict” state so that other threads can safely evict the cache-line. This approach effectively reduces cache thrashing and improves the performance of GNN feature aggregation on GPUs.

\subsection{Graph Structure Data in CPU Memory}

\begin{figure}[h]
    \centering
    \includegraphics[width=\columnwidth]{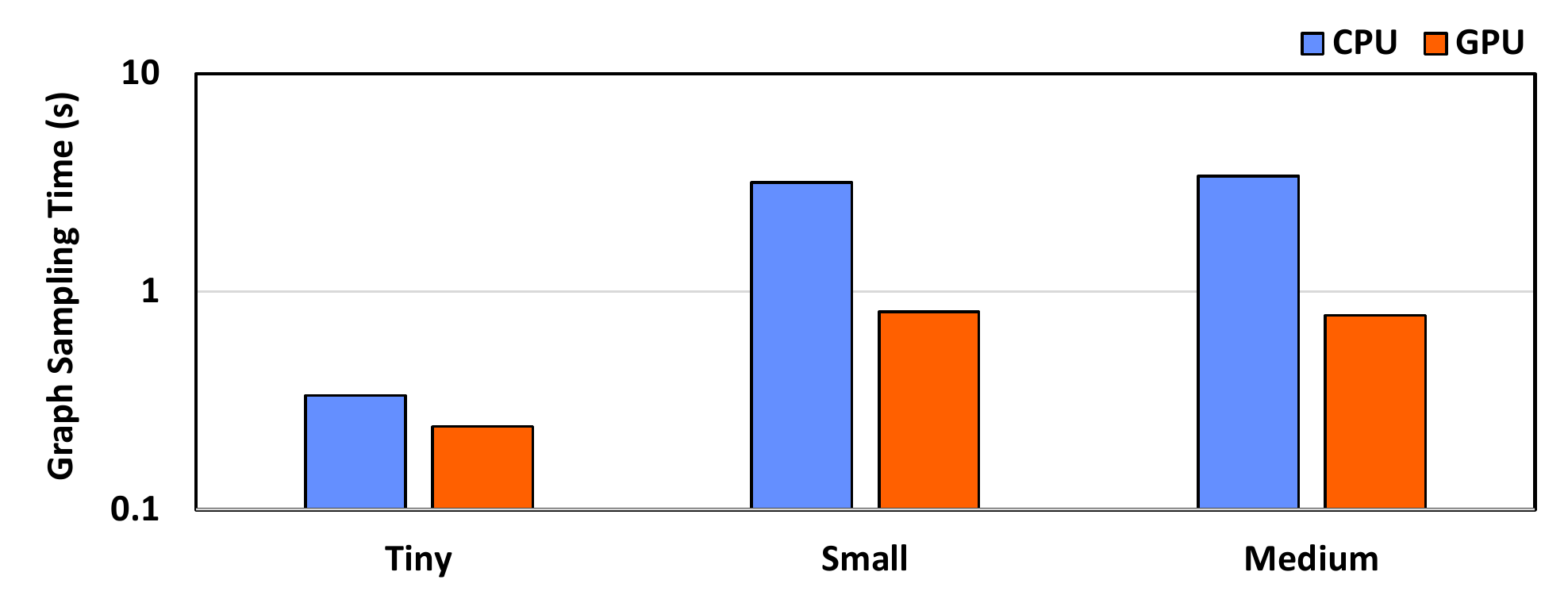}
    \caption{Graph sampling time of CPU and GPU graph sampling on the graphs with different sizes }
    \label{fig:sampling_eval}
\end{figure}

As shown in Figure~\ref{fig:breakdown}, the graph sampling throughput is higher on GPU than on CPU despite graph sampling being a sequential process.
This is because the graph sampling process is especially latency-critical for large-scale graphs. The fundamental approach to accelerate such a process is to exploit parallelism to hide the latency, which GPUs naturally provide. Figure~\ref{fig:sampling_eval} shows that GPU outperforms CPU for all three datasets, with a performance gain of over 3$\times$ for the medium dataset. 
However, storing graph structure data in storage incurs multiple problems.

Firstly, the graph sampling process has a smaller data access granularity than the feature aggregation process, resulting in significant I/O amplification. This is because the data accesses to the storage devices are handled in page granularity, such as 4KB, meaning even if only a small segment of data is requested, the entire cache-line is transferred from the storage to GPU memory. Secondly, the random data access pattern from the sampling process makes it challenging for the GPU cache to exploit data locality, which can degrade the performance of the feature aggregation process. This is because the GPU memory is a limited resource, and the random data access pattern can pollute the GPU software-defined cache.

To address these challenges, 
the \pname{} dataloader employs zero-copy data transfer via Unified Virtual Addressing (UVA) for graph structure data. Instead of storing the entire graph data in storage devices, our dataloader allows users to store node feature data on storage while pinning graph structure data in the CPU memory. This makes it possible to execute the graph sampling process on either CPU or GPU. 
This is a practical approach because the graph structure data is small compared to the node feature data, even for the terabyte-scale graphs that we expect to accommodate in the foreseeable future, as shown in Table~\ref{tab:dataset_percent}.

\section{Evaluation}

\subsection{Experimental Setup}

\textbf{Environment.} Table~\ref{tab:config} summarizes the system configuration for all evaluations. We compare \pname{} and the state-of-the-art baseline dataloaders on an AMD EPYC high-end server-grade system equipped with a NVIDIA A100-40GB GPU and 1TB DDR4 CPU DRAM. 
Additionally, either 768 GB or 512GB of the CPU memory was locked for exclusion to limit the CPU memory capacity for evaluation purposes.
The evaluations were conducted using Intel Optane PCIe Gen4 NVMe SSDs as the default storage. To comprehensively assess overall performance, measurements were also taken with Samsung 980 Pro SSDs.

{\renewcommand{\arraystretch}{1.2}
\begin{table}[h]
\centering
\scriptsize
    \caption{\small {Configuration used to evaluate \pname{}.}}
    \begin{tabular}{|p{0.55in}|p{2.25in}|}
    \hline
    {\textbf{Configuration}}& {\textbf{Specification}} \\
    \hline
    \hline
	{CPU}                  & {AMD EPYC 7702 64-Core Processor} \\
    \hline
	{Memory}               & {1TB DDR4} \\
    \hline
	\multirow{3}{*}{GPU} &  NVIDIA A100 HBM2 40GB \\ 
	                       &  108 SMs, 192KB Shared Memory per SM \\
	                       &  40MB LLC, 1555GBps HBM Bandwidth  \\
    \hline
    \multirow{4}{*}{S/W}   & Ubuntu 20.04 LTS, NVIDIA Driver 470.103 \\
                           & CUDA 11.4 \\
                           & DGL 0.10 \\
                           & Pytorch 1.13.0 \\
    \hline
    \multirow{2}{*}{SSDs}     &   Intel Optane SSDs \\
                            & Samsung 980 Pro SSDs \\
                             &   PCIe Gen 4 Interconnect \\
    \hline
\end{tabular}
\label{tab:config}
\end{table}
}

\textbf{Datasets.} To assess the performance of \pname{} dataloader on large-scale graph datasets, we conducted experiments using four real-world datasets: IGB-Full~\cite{IGB}, IGBH-Full~\cite{IGB}, ogbn-papers100M~\cite{ogbn_paper}, and MAG240M~\cite{MAG}. Table~\ref{tab:dataset} presents the characteristics of these datasets, such as the number of nodes and edges, the dimension of the node feature data, and the type of graph. It is worth noting that ogbn-papers100M and MAG240M datasets are small enough to fit into the CPU memory of our evaluation system.

{\renewcommand{\arraystretch}{1.2}
\begin{table}[h]
\centering
\scriptsize
    \caption{\small Real-world dataset used for evaluating \pname{}. 
}

\begin{tabular}{|p{0.6in}|p{0.6in}|>{\raggedleft\arraybackslash}p{0.5in}|>{\raggedleft\arraybackslash}p{0.5in}|>{\raggedleft\arraybackslash}p{0.40in}|}
    \hline
    \textbf{Dataset} & \textbf{Graph Type} & \textbf{Number of Nodes}& \textbf{Number of Edges}& \textbf{Feature Dimension}  \\
    \hline
    \hline
	ogbn-papers100M & Homogeneous & 111,059,956  & 1,615,685,872 & 128 \\
    \hline
	IGB-Full & Homogeneous &  269,364,174 & 3,995,777,033 &  1024 \\
    \hline
	MAG240M & Heterogeneous & 244,160,499 & 1,728,364,232 &  768 \\
    \hline
	IGBH-Full & Heterogeneous & 547,306,935 & 5,812,005,639 & 1024  \\
    \hline
\end{tabular}
\label{tab:dataset}
\end{table}
}

{\renewcommand{\arraystretch}{1.2}
\begin{table}[h]
\centering
\scriptsize
    \caption{\small IGB datasets used for micro-benchmarks.
}
\begin{tabular}{|p{0.6in}|p{0.6in}|>{\raggedleft\arraybackslash}p{0.5in}|>{\raggedleft\arraybackslash}p{0.5in}|>{\raggedleft\arraybackslash}p{0.40in}|}
    \hline
    \textbf{Dataset} & \textbf{Graph Type} & \textbf{Number of Nodes}& \textbf{Number of Edges}& \textbf{Feature Dimension}  \\
    \hline
    \hline
	IGB-tiny & Homogeneous & 100,000  & 547,416 & 1024  \\
    \hline 
    IGB-small & Homogeneous & 1,000,000  & 12,070,502 & 1024 \\
    \hline
	IGB-medium & Homogeneous & 10,000,000 & 120,077,694 &  1024 \\
    \hline
	IGB-large & Homogeneous & 100,000,000 & 1,223,571,364 &  1024 \\
    \hline
\end{tabular}
\label{tab:IGBdataset}
\end{table}
}

{\renewcommand{\arraystretch}{1.2}
\begin{table}[h]
\centering
\scriptsize
    \caption{\small Datasize distribution for the real-world datasets.
}

\begin{tabular}{|p{0.4in}|>{\raggedleft\arraybackslash}p{0.7in}|>{\raggedleft\arraybackslash}p{0.7in}|>{\raggedleft\arraybackslash}p{0.7in}|}
    \hline
    \textbf{Dataset} & \textbf{Feature Data Size (\%)}& \textbf{Graph Structure Data Size (\%)}& \textbf{Total Size (GB)}  \\
    \hline
    \hline
	ogbn-papers100M & 68.3 & 31.0  &  77.4 \\
    \hline
   
	IGB-Full &   94.7 & 5.1 & 1084.0 \\
    \hline
	MAG240M & 86.7 & 12.8 & 200.0  \\
    \hline
	IGBH-Full & 96.0 & 3.8 &  2773.0  \\
    \hline
\end{tabular}
\label{tab:dataset_percent}
\end{table}
}

\textbf{\pname{} Implementation} We extended DGL~\cite{dgl} to implement the \pname{} dataloader. Our approach involves creating new extensions for the storage-based feature gathering by leveraging BaM~\cite{bam} to support user-level GPU-initiated direct storage access. We then extended the DGL dataloader class to incorporate \pname{} functionalities. To use the \pname{} dataloader, users only need to set the \pname{} flag when initializing the DGL dataloader.

\textbf{Model:} We assessed the performance of the \pname{} dataloader using two distinct sampling techniques: neighborhood sampling~\cite{Graphsage} and LADIES~\cite{LADIES} for layer-wise sampling. All models were configured with a hidden dimension of 128, and a mini-batch size of 4,096 was employed with three sampling layers

\textbf{GIDS Dataloader:} In the default configuration, we allocated 8 GB of GPU device memory for GPU software-defined caching and allocated CPU memory for 10\% of the dataset. We utilized a single NVMe SSD for both the \pname{} dataloader and the DGL baseline dataloader.

\textbf{Baseline:} We compared \pname{} with the DGL dataloader that is extended to work with memory-mapped files. We used the memmap function from NumPy to create a memory-mapped array tensor for the graph data. 
Additionally, we implemented a BaM dataloader, which integrates the BaM system into the DGL dataloader, and compared it with \pname{} to showcase the novel benefits offered by \pname{}. 
Furthermore, we conducted comparisons with Ginex~\cite{Ginex}. However, Ginex exclusively supports homogeneous graphs and neighborhood sampling techniques. 

\textbf{Measuring Execution Time:} When working with large graph datasets, the training process can be excessively long, especially for the baseline. Therefore, we conducted the evaluations by measuring the execution time for 100 iterations after a warm-up stage of 1,000 iterations. We used the listed model configuration, with a mini-batch size typically ranging from 1 GB to 3 GB. This setup is 
favorable for the baselines mmap and Ginex as 
we are not measuring the storage latency overhead from the first 1,000 iterations when the page cache in the CPU memory is being warmed up for the baseline. 
However, for the \pname{} dataloader, only 10 iterations are required to warm up the GPU software-defined cache, and the cache miss for the baseline is more critical due to the exposed storage latency.

\subsection{Estimation of the Required Number of Overlapping Storage Accesses}

In Section~\ref{sec:ssd}, we introduced a mathematical model to estimate the necessary number of overlapping storage accesses to attain the target SSD bandwidth. To validate our model, we measured the achieved SSD bandwidth with different numbers of overlapping storage accesses for two distinct SSDs: the Intel Optane SSD and the Samsung 980 Pro SSD. Then, we compared the achieved SSD bandwidth with the expected SSD bandwidth calculated by our model. 
For this evaluation, we configured our IO size to be 4KB. With a 4KB IO size, the peak IOPs reached 1.5M IOPs for Intel Optane and 700K IOPs for the Samsung 980 Pro SSDs. The SSD latency was measured at 11$\mu$s for Intel Optane SSDs and 324$\mu$s for Samsung 980 Pro SSDs. Additionally, we added 25$\mu$s to account for the initial latency related to kernel launch and to capture initial software overheads while we set the termination latency to 5$\mu$s.

\begin{figure}[h]
\centering
\includegraphics[width=\columnwidth]{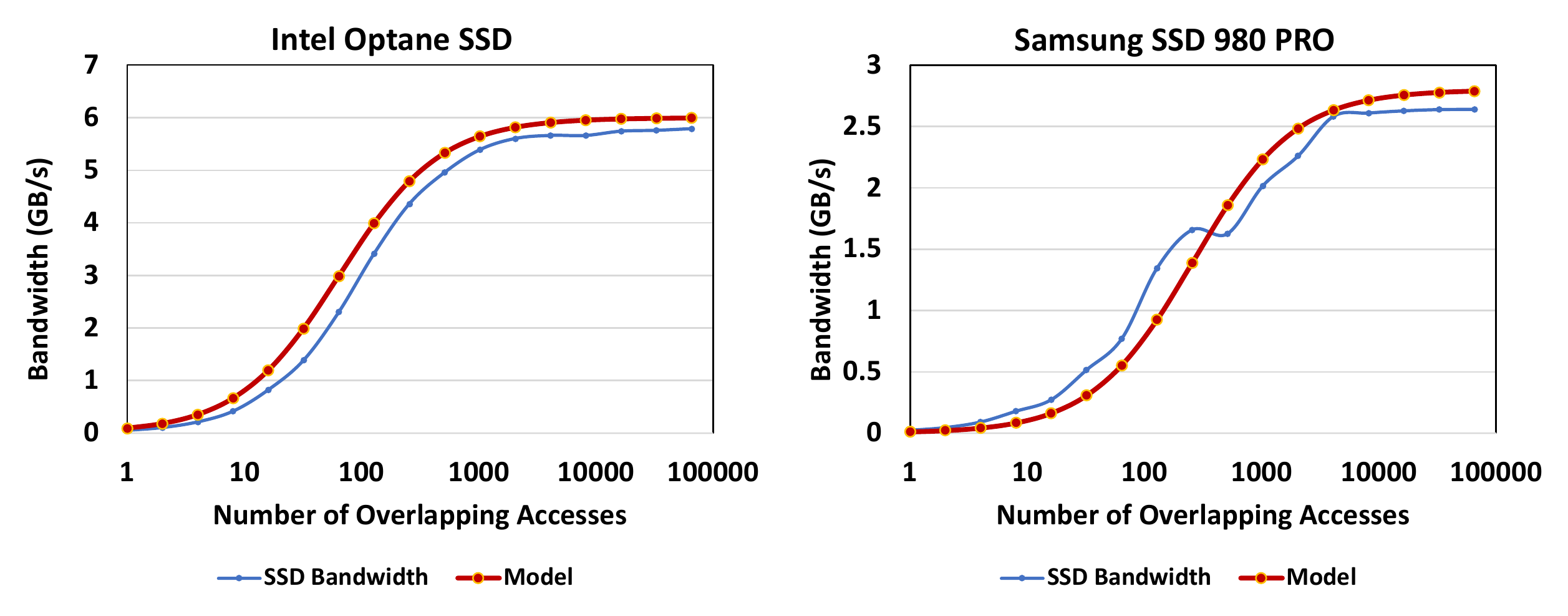}
\caption{SSD bandwidth with different numbers of overlapping accesses using estimation from the model and measurement from the microbenchmark. The model accurately predicts both the trend and values of the measurements.}
\label{fig:latency_model}
\end{figure}

Figure~\ref{fig:latency_model} displays the achieved SSD bandwidth and the expected SSD bandwidth from our model based on the number of overlapping SSD accesses. Despite the high variance in latency, our model accurately estimates the SSD bandwidth, particularly when it approaches the peak bandwidth. For example, if we aim to achieve 95\% of the peak SSD IOPs, our model estimates that 812 accesses are required for the Intel Optane SSD, while we measured the targeted IOPs with 1024 overlapping accesses.
These results show that our model accurately estimates the required number of overlapping storage accesses for \pname{} techniques, such as the dynamic storage access accumulator.

\subsection{Impact of the Dynamic Storage Access Accumulator}

In this evaluation, we present an experiment using two Intel Optane SSDs connected to a single GPU to assess the effectiveness of the dynamic storage access accumulator. We varied the batch size across a range from 32 to 128 while keeping the fan-out values constant for neighborhood sampling at (5,5). Our evaluation employed the IGB-Full dataset, and we measured the GPU PCIe ingress bandwidth.

\begin{figure}[h]
\centering
\includegraphics[width=\columnwidth]{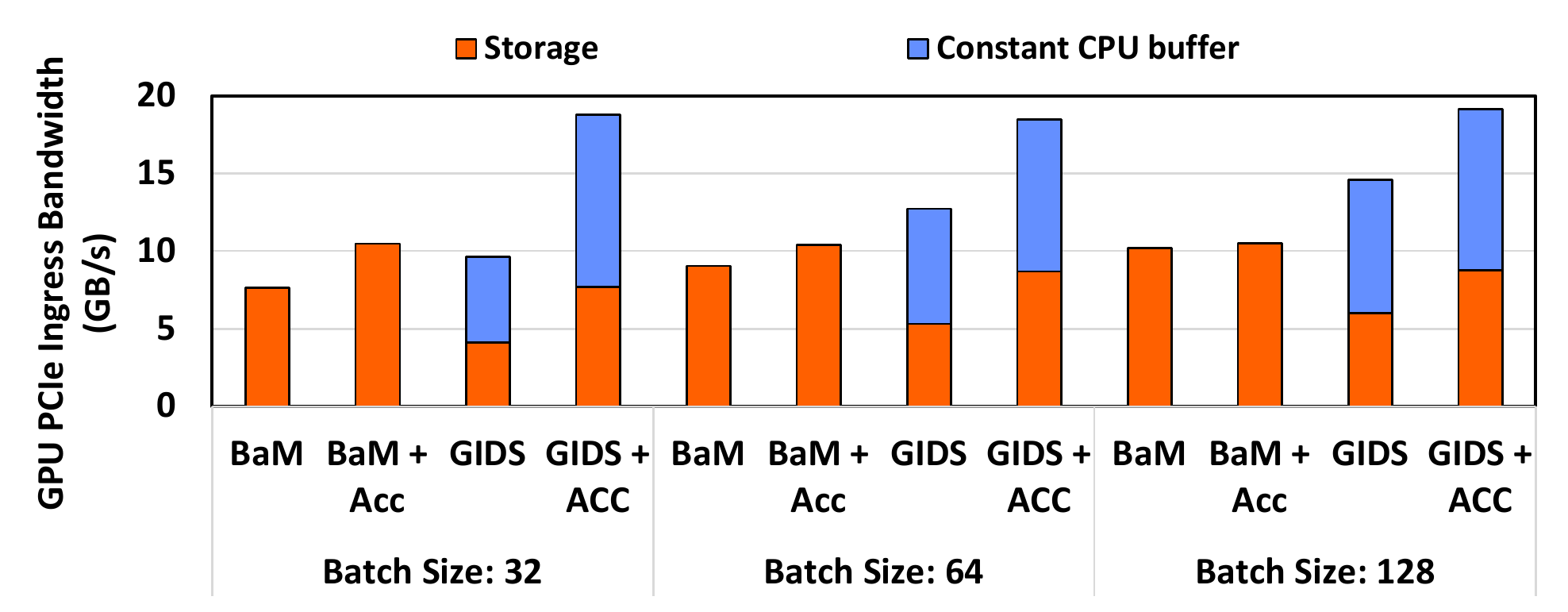}
\caption{The dynamic storage access accumulator increases GPU PCIe ingress bandwidth for BaM (1.25$\times$) and \pname{} dataloader (1.95$\times$) by enhancing overlapping storage accesses. This improvement is more pronounced in \pname{} due to its lower SSD bandwidth use through (1) a higher software cache hit ratio due to window buffering and (2) a reduced number of storage accesses through the CPU constant buffer.
}

\label{fig:stroage_access_accumulator}
\end{figure}

Figure~\ref{fig:stroage_access_accumulator} illustrates the GPU PCIe ingress bandwidth with different configurations during the feature aggregation stage of the \pname{} dataloader. The BaM dataloader integrates the BaM system into the DGL dataloader while window buffering and the constant CPU buffer are activated for the \pname{} dataloader.

The baseline BaM dataloader achieves PCIe ingress bandwidths of 7.6 GB/s, 9.4 GB/s, and 10.1 GB/s for batch sizes of 32, 64, and 128, respectively. Since the peak bandwidth for Intel Optane SSDs is approximately 5.8 GB/s, the peak collective SSD bandwidth is 11.6 GB/s, showing that there are insufficient overlapping storage accesses to effectively hide latency, particularly evident with a batch size of 32. With the accumulator, BaM can achieve 9.8 GB/s, 10.4 GB/s, and 10.6 GB/s, which is much closer to the peak bandwidth. 

With the incorporation of the constant CPU buffer and window buffering, the performance gap widens. \pname{} with the accumulator achieves 1.95$\times$, 1.46$\times$, and 1.31$\times$ speedup compared to \pname{} without the accumulator. The number of concurrent storage accesses is reduced in \pname{} as some storage requests are redirected to either the GPU software cache or the constant CPU buffer, resulting in lower SSD bandwidth utilization. Thus, the performance gain by the accumulator is higher in \pname{} as it ensures the peak SSD bandwidth utilization even with redirected storage access. While the achieved SSD bandwidth is slightly below the peak, this is due to a decrease in the number of GPU threads that can simultaneously enqueue storage accesses, as they are involved in copying data from the CPU buffer to GPU memory.
Overall, the dynamic storage access accumulator empowers users to enhance SSD bandwidth utilization for various configurations of their GNN models, irrespective of their hardware specifications.

\subsection{Impact of the Constant CPU Buffer}

In this section, we 
examine the impact of the constant CPU buffer on the feature aggregation performance, particularly when the PCIe bandwidth is underutilized. In this evaluation, the \pname{} dataloader fetches feature data from storage that consists of a single SSD for the IGB-full dataset.

Across all dataloaders, we consistently configured the GPU software cache to be 8GB, without the application of the window buffering technique. To assess the influence of the CPU buffer, we systematically varied its size, ranging from 10\% to 20\% of the dataset size.
Furthermore, we explored the performance of feature aggregation when implementing the reverse page-rank algorithm to determine which nodes should be pinned in the constant CPU buffer.

\begin{figure}[h]
    \centering
    \includegraphics[width=\columnwidth]{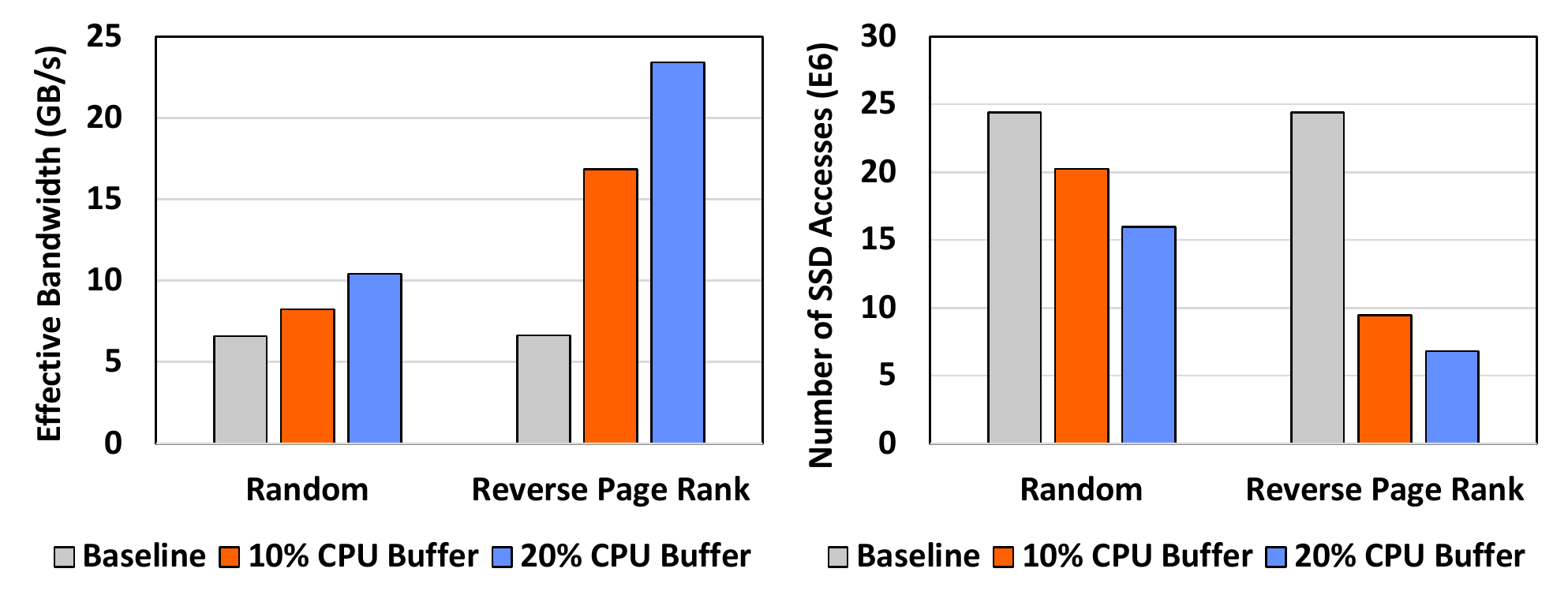}
    \caption{Feature aggregation throughput of the baseline \pname{} and \pname{} with the constant CPU buffer. \pname{} achieves up to 3.53$\times$ higher effective bandwidth with the constant CPU buffer with reverse page rank. With reverse page rank, the CPU constant buffer holding 20\% of the graph feature data effectively magnifies the bandwidth of a single SSD to that of four SSDs.}
    \label{fig:cpu_buffer}
\end{figure}

Figure~\ref{fig:cpu_buffer} provides insights into the effective bandwidth when 10\% or 20\% of the feature data is pinned in the constant CPU buffer. The baseline \pname{} dataloader achieved a feature aggregation bandwidth of 6.6 GBps, which slightly exceeds the peak SSD bandwidth (5.8 GBps) as it fully saturated the SSD bandwidth, and some accesses were redirected to the GPU software cache. However, it is essential to highlight that the baseline dataloader cannot fully utilize the available GPU ingress PCIe bandwidth.


In contrast, with a 20\%  constant CPU buffer size, particularly when employing the reverse page-rank selection strategy, the feature aggregation throughput of the \pname{} dataloader is increased from 10.4 GBps and 23.4 GBps. This is because a significant portion of storage accesses are redirected to the constant CPU buffer, increasing the PCIe bandwidth utilization beyond the SSD bandwidth. 
These results show that \pname{} dataloader's capacity to mitigate resource constraints on SSDs by harnessing the potential of CPU resources, establishing it as a practical solution across a wide spectrum of systems.

\subsection{Impact of the Window Buffering Cache Optimization}
\label{sec:gpu_optimiazation}

In this section, we present an evaluation of the impact of GPU software-defined cache optimization on the feature aggregation process. 
To conduct this evaluation, we compared the performance of \pname{} with a basic GPU software-defined cache against \pname{} with window buffering optimization. 
To ensure a fair comparison, we used the IGB-full dataset with the same neighborhood sampling parameters and mini-batch size. We evaluated the performance with {an} 8 GB GPU software cache.

\begin{figure}[h]
    \centering
    \includegraphics[width=\columnwidth]{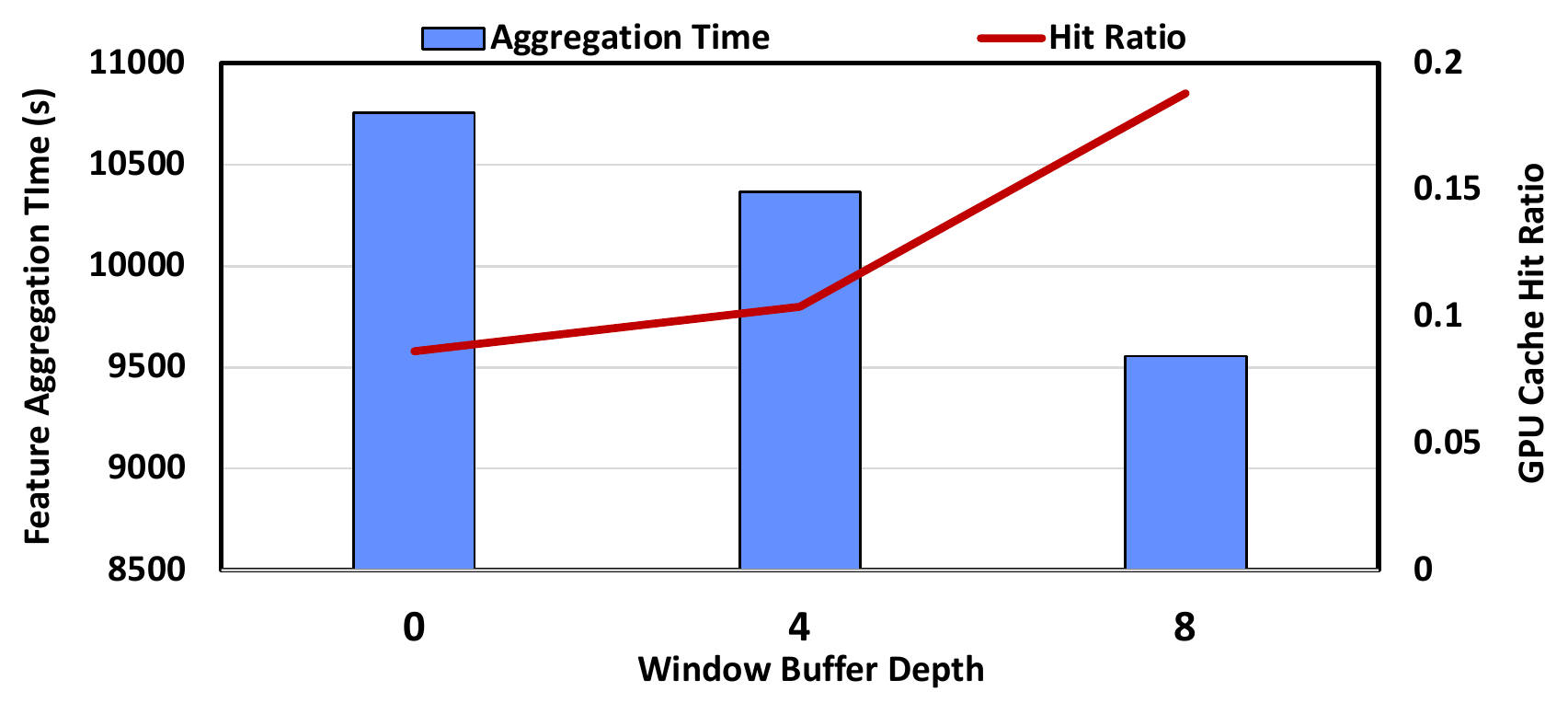}
    \caption{Performance comparison of feature aggregation process on \pname{} dataloader for different window buffering depths.}
    \label{fig:window_buffer_eval}
\end{figure}

To accurately measure the impact of the window buffering technique, we varied the depth of the window buffer from 0 to 4, and then to 8 while evaluating the feature aggregation time and the GPU software-defined cache hit ratio. When the window buffer depth is 0, the GPU software-defined cache follows the random eviction policy, which serves as the baseline. Figure~\ref{fig:window_buffer_eval} displays the results, which show that the window buffering technique can improve the cache hit ratio. A window size of 4 improves the cache hit ratio by only 1.2$\times$ and the feature aggregation time by 1.04$\times$.

Setting the window buffer depth too low, compared to the size of the GPU cache, can lead to a similar performance as random eviction. For instance, if the mini-batch size is 2 GB, and the GPU cache size is 10 GB, most of the node features from the previous four mini-batches still reside in the cache with a random eviction policy. Therefore, the optimal hit ratio with a window size of four is similar to random eviction, making it hard to achieve a meaningful performance gain.

When we increase the window buffer size to 8, the cache hit ratio improves by 2.19$\times$ over not having any window buffering, and the aggregation time decreases by 1.13$\times$. This is because the depth of the window buffer provides enough information about the cached node features that will be reused in future mini-batches to avoid evicting reusable cache-lines across mini-batches, which results in a substantial difference compared to random eviction. When the window buffer size is set to 8, the cached node features that the GPU cache exceeds the number of the node features that can fit into the GPU cache. Any further increase in the window buffer depth should be accompanied by an increased GPU cache size.

\begin{figure}[h]
    \centering
    \includegraphics[width=\columnwidth]{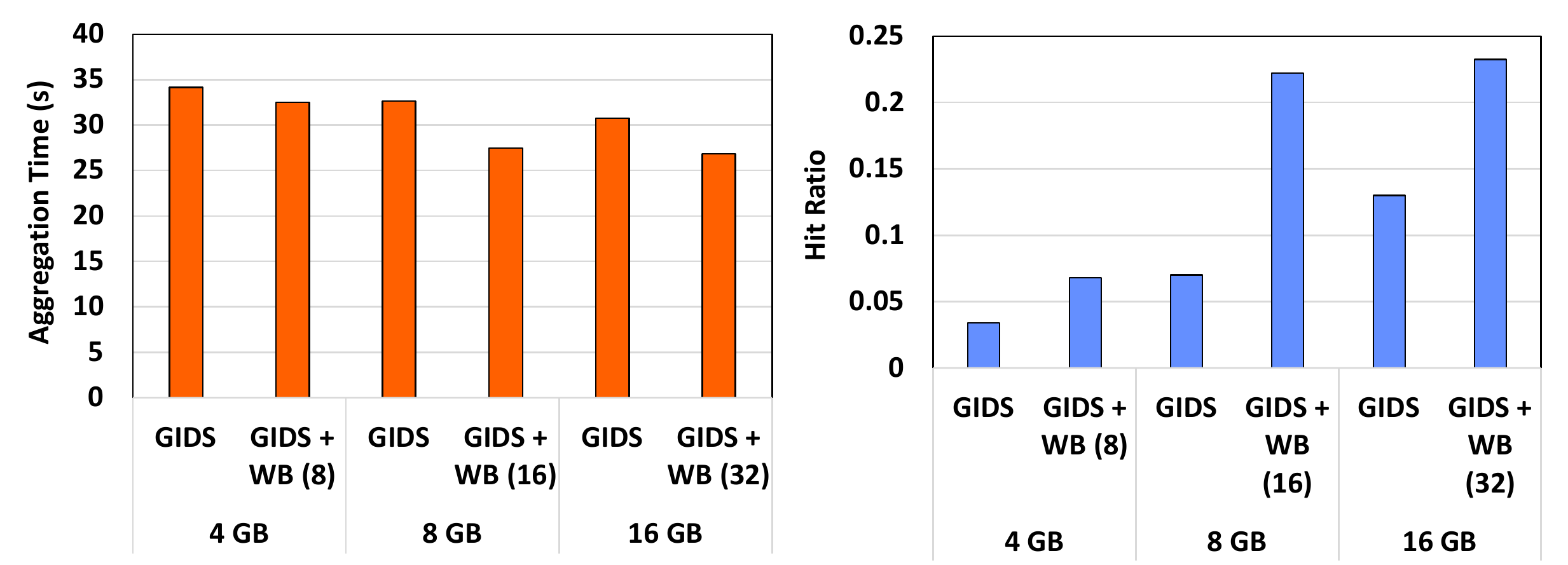}
    \caption{Feature aggregation performance comparison between window buffering and the baseline. \pname{} achieves a higher GPU software cache hit ratio by capturing more locality with the window buffer.}
    \label{fig:window_pin_comparison}
\end{figure}

Next, we compare the performance of the window buffering technique with 4 GB, 8 GB, and 16 GB of the GPU software cache. When window buffering is activated, we set the depth of the window buffer to 16 for all 4 GB, 8 GB, and 16 GB GPU software cache, respectively.

As shown in Figure~\ref{fig:window_pin_comparison}, the window buffering technique demonstrates an improvement, outperforming the GIDS without window buffering by a factor of 1.20, 1.18, and 1.12 for the IGB-Full dataset with 4 GB, 8 GB, and 16 GB GPU cache, respectively. The hit ratio for the baseline GPU software cache increases as the size of the cache increases since it can exploit more temporal locality. However, even the 16 GB GPU cache performs worse than the 4 GB GPU cache with window buffering because the hit ratio of \pname{}'s GPU cache is less affected by the GPU cache size unless the window buffer depth is changed.

However, there is a trade-off to consider when increasing the window buffer depth. First, there needs to be enough memory space for the window buffer. As the number of node samples for each mini-batch is around 1M, the size of the list of sampled nodes for a mini-batch is several megabytes. 
Although this is not a significantly large amount, larger window sizes increase the GPU memory requirement as the list of sampled nodes in the window buffer must be kept in GPU memory for subsequent iterations.
Additionally, a larger window size means a larger portion of the GPU cache will be pinned for future reuse, increasing the contention on the available cache-lines in the GPU software cache. 
Therefore, it is essential to carefully choose the window buffer size to ensure that the benefit of a higher cache hit ratio outweighs the overhead of a larger window buffer size.
By default, the \pname{} dataloader sets the depth of the window buffer to 8 based on the system environment. 
However, the window buffer depth is a tunable parameter that users can adjust based on the hardware environment, such as GPU memory size.

\begin{figure*}[h]
    \center
    \includegraphics[width=\textwidth]{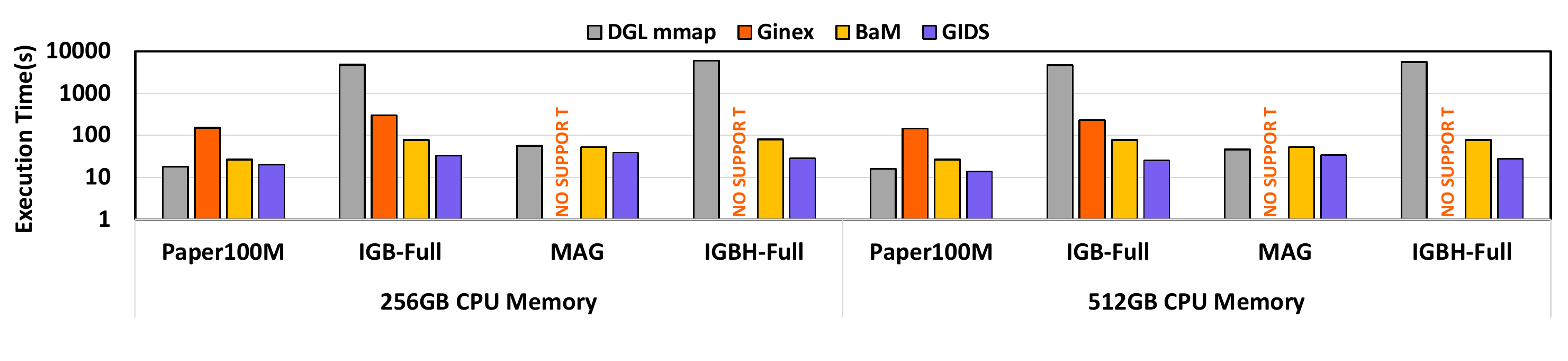}
    \caption{End-to-End (E2E) GNN training performance comparison of the \pname{} dataloaders and the baseline dataloaders with Samsung  980pro SSDs.  \pname{} achieves up to 582$\times$, 	10.62$\times$, and 3.09$\times$ speed up compared to DGL mmap, Ginex, and BaM dataloaders, respectively.}
    \label{fig:overall_eval_ss}
\end{figure*}

\begin{figure*}[h]
    \center
    \includegraphics[width=\textwidth]{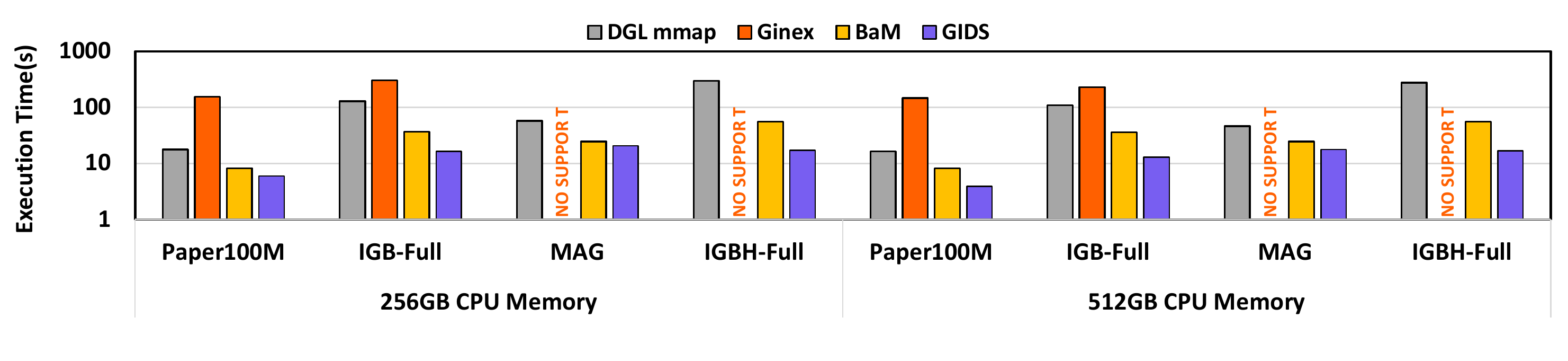}
    \caption{End-to-End (E2E) GNN training performance comparison of the \pname{} dataloaders and the baseline dataloaders with Intel Optane SSDs. \pname{} achieves up to 17.28$\times$, 	37.21$\times$, and 3.23$\times$ speed up compared to DGL mmap, Ginex, and BaM dataloaders, respectively.}
    \label{fig:overall_eval_I}
\end{figure*}

\subsection{Overall Performance}

Figure~\ref{fig:overall_eval_ss} and Figure~\ref{fig:overall_eval_I} illustrate the End-to-End (E2E) GNN training times for both the baseline and \pname{} dataloaders on homogeneous and heterogeneous graphs, using Samsung 980pro and Intel Optane SSDs, respectively. Notably, Ginex does not support heterogeneous graphs, and therefore, the performance on IGBH-Full and MAG240M datasets for Ginex is not measured. For IGBH-Full datasets, two SSDs are used for the evaluation due to storage capacity.

As shown in Figure~\ref{fig:overall_eval_ss} and Figure~\ref{fig:overall_eval_I}, the \pname{} dataloader achieves speedups, reaching up to 8.3$\times$ and 582$\times$ compared to the DGL baseline dataloader for Intel Optane and Samsung 980pro SSDs, respectively. The performance gain is higher with Samsung 980pro SSDs, primarily because the feature aggregation process in the baseline dataloader is limited by the SSD latency, and the SSD read latency of Samsung 980pro SSDs is approximately 30$\times$ higher than that of Intel Optane SSDs. Furthermore, the performance gain for IGB-Full and IGBH-Full datasets is substantially larger than that for ogbn-papers100M and MAG240M because the sizes of the latter two graphs are smaller than the CPU memory capacity, and thus the baseline does not incur a significant number of page faults while training with these datasets.

When compared with Ginex, \pname{} attains speedups of up to 10.6$\times$ and 37.2$\times$ with Intel Optane and Samsung 980pro SSDs. Ginex aims to alleviate storage overhead by reducing redundant accesses to storage, but the storage latency remains a challenge to effectively mitigate, resulting in significant overhead. Finally, \pname{} also outperforms the BaM dataloader by 1.3$\times$ to 3.1$\times$. This is attributed to \pname{}'s efficient utilization of CPU and GPU memory, which minimizes storage accesses and leads to higher effective bandwidth.

\begin{figure}[h]
    \center
    \includegraphics[width=\columnwidth]{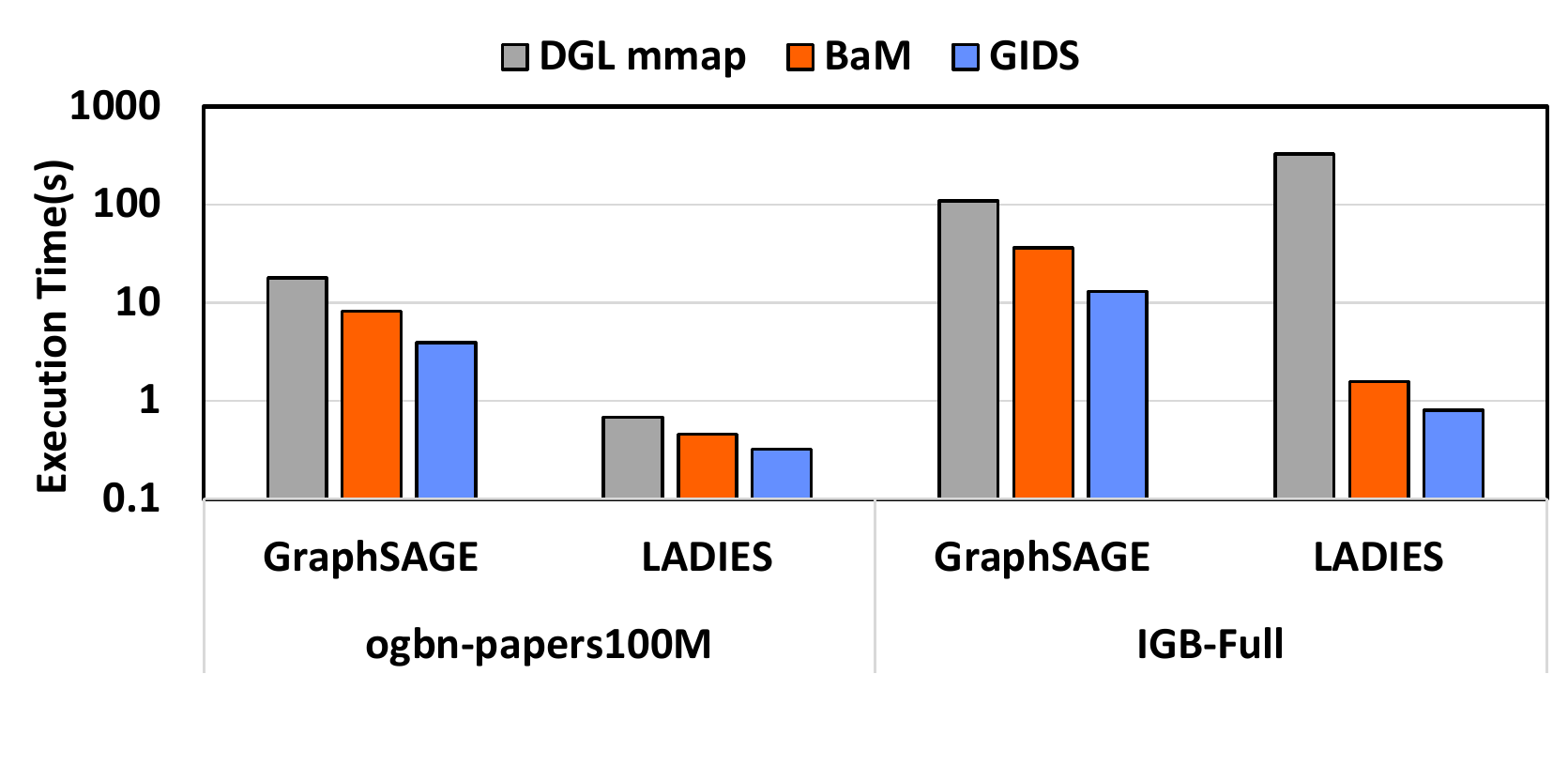}
    \caption{Feature aggregation time comparison of the \pname{} dataloaders and the baseline dataloaders for neighborhood and LADIES sampling.}
    \label{fig:overall_ladies}
\end{figure}

\subsection{Performance of \pname{} with Layer-wise Sampling}

We also conducted a performance comparison of \pname{} with layer-wise sampling techniques, such as LADIES~\cite{LADIES}, against the baseline dataloaders. Since Ginex~\cite{Ginex} does not support sampling techniques other than neighborhood sampling, we compared \pname{} with the DGL dataloader and BaM. In this evaluation, we pinned 512 GB of CPU memory while allocating 8 GB GPU cache for both BaM and \pname{}. As shown in Figure~\ref{fig:overall_ladies}, \pname{} achieved a speedup of 412$\times$ compared to the DGL dataloader and a 1.92$\times$ speedup compared to BaM. These results highlight \pname{}'s exceptional performance with layer-wise sampling techniques.

For subgraph-based sampling techniques, such as ClusterGCN~\cite{ClusterGCN}, the \pname{} dataloader can also be utilized. 
However, subgraph-based sampling techniques involve the use of the Metis~\cite{Metis} algorithm to partition the graph and feature vectors to fit in the CPU memory. Metis-based graph dataset partition is an extremely time-consuming process for large-scale graph datasets like IGB (more than 2 days). On the other hand, GIDS leverages SSDs to store graph datasets and enables the mapping of significantly large graph datasets in a single node (depending on the size and number of SSDs in the system) without the need for a graph partitioning step. Given this, we chose not to evaluate \pname{} for subgraph-based partitioning due to the potential impracticality of employing the Metis algorithm in such cases.

\section{Related Work}
\label{sec:related}

Several GNN specific applications and optimizations have been proposed in the literature~\cite{related1,related2,related3,related4,related5,related6}. ROC~\cite{ROC}, NeuGraph~\cite{Neugraph}, and DSP~\cite{DSP} propose multi-GPU training system for large-scale GNN training. However, they require significant additional hardware resources and are not scalable solutions.

FeatGraph~\cite{featgraph} and ZIPPER~\cite{zipper} propose tiling to mitigate the memory footprint during GNN training. FeatGraph reduces memory usage by utilizing graph partitioning and feature dimension tiling. Meanwhile, ZIPPER employs graph-native intermediate representation to optimize GNN, such as sparse graph tiling and redundant operation elimination. However, these approaches suffer from random accesses from GNN, leading to poor performance. Moreover, these solutions do not leverage GPU for the data preparation process.

AliGraph~\cite{AliGraph}, PaGraph~\cite{PaGraph}, and Ginex~\cite{Ginex} use in-memory caching to reduce data transfer overhead. AliGraph and PaGraph cache high out-degree vertices in GPU memory to minimize data transfer between CPU and GPU. Ginex uses Belady’s algorithm with super-batch samples and pipelining techniques to hide the latency from specialized caching policies. However, these approaches rely on the CPU for the data preparation process and cannot fully hide storage latency.

Data Tiering~\cite{DataTiering} uses weighted reverse PageRank to estimate the frequency of accesses during node sampling, improving GPU memory utilization. However, it requires all graph data to be stored in either CPU or GPU for GNN training execution, so it is not applicable to large-scale GNN training. 




\section{Conclusion}

Training Graph Neural Networks (GNNs) on large-scale graph datasets is a challenging task due to their size exceeding the CPU memory capacity. Although distributed training is a possible solution, it is not cost-effective or even practical for many users. In this paper, we propose the \pname{} dataloader, a GPU-oriented GNN training system that enables the training of large-scale graph datasets on a single machine. 
The \pname{} dataloader enables GPU threads to directly access storage and fully tolerates the long storage latency by exploiting the massive data-level parallelism provided by GPUs and our novel storage access accumulator. Moreover, the \pname{} dataloader further improves performance by utilizing GPU memory as a software-defined cache with window buffering and CPU memory as the constant CPU buffer. By reducing the I/O overhead and maximizing hardware resource utilization, \pname{} dataloader can scale GNN training to datasets whose sizes are more than an order of magnitude larger than a single machine's CPU memory capacity while achieving up to 582$\times$ speedups over the state-of-the-art dataloader for the overall execution of an end-to-end GNN training pipeline.

\begin{acks}

We would like to acknowledge all of the help from members of
the IMPACT research group, the IBM-Illinois Center for Cognitive
Computing Systems Research (C3SR) and NVIDIA Research without which we could not have achieved the results reported in this paper. 
Special thanks to Kun Wu, Isaac Gelado, and Scott Mahlke who generously shared their insights through numerous discussions. 
This work uses GPUs donated by NVIDIA
and is partially supported by the IBM-ILLINOIS C3SR and by the
IBM-ILLINOIS Discovery Accelerator Institute (IIDA).

\end{acks}


\bibliographystyle{ACM-Reference-Format}
\bibliography{sample-base}

\end{document}